# Realization of *p*-valued Toffoli/Deutsch quantum gates with two and three controls, and mixed polarity

**Claudio Moraga**

# Realization of *p*-valued Toffoli/Deutsch quantum gates with two and three controls, and mixed polarity

*Abstract* **In this report reversible Toffoli and quantum Deutsch gates are extended to the *p*-valued domain. Their structural parameters are determined and their behavior is proven. Both conjunctive and disjunctive control strategies with positive and mixed polarities are introduced for the first time in a *p*-valued domain. The design is based on elementary Muthukrishnan-Stroud quantum gates, hence the realizability of the extended gates in the context of ion traps should be possible.**

*Keywords* **Toffoli-Deutsch gates, conjunctive-disjunctive control, mixed polarity control, *p*-valued quantum computing.**

## I Introduction

Toffoli gates [22] gave a considerable impulse to the development of reversible binary computing, because of their simplicity and for being functionally complete. Toffoli introduced the gate as “controlled-controlled-Not” (*CCN*), but soon started to be called “Toffoli gate”. Its symbolic representation is shown in Fig. 1.

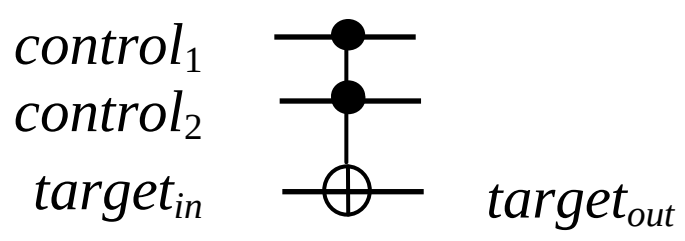

Fig. 1 Symbol of the Toffoli gate

In the symbolic representation of Fig. 1, the *control* signals are passed to the output without change.

The functionality of the gate may be expressed in three different (equivalent) ways:

i) The *CCN* view.

$$\begin{aligned} target_{out} &= \text{Not}(target_{in}) \quad \text{iff} \quad control_1 = control_2 = 1 \\ &= target_{in} \qquad \text{otherwise.} \end{aligned} \tag{1}$$

ii) The EXOR view

$$target_{out} = target_{in} \oplus \text{AND}(control_1, control_2),$$

leading to

$$\begin{aligned} &target_{out} = target_{in} \oplus 1 \text{ iff } control_1 = control_2 = 1, \text{ i.e.} \\ &target_{out} = \text{Not}(target_{in}) \end{aligned} \tag{2}$$

iii) The matrix view. Let $\boldsymbol{T}$ specify the transfer function of the gate. In $\boldsymbol{T}$, the rows are ordered lexicographically according to $control_1$, $control_2$, $target_{in}$ and the columns, according to $control_1$, $control_2$, $target_{out}$. This is shown in (3), where 2×2 identity submatrices are shaded whenever at least one control signal is 0 and a 2×2 inverting submatrix when both control signals equal 1.

$$
\boldsymbol{T} = \begin{bmatrix} 1 & 0 & 0 & 0 & 0 & 0 & 0 & 0 \\ 0 & 1 & 0 & 0 & 0 & 0 & 0 & 0 \\ 0 & 0 & 1 & 0 & 0 & 0 & 0 & 0 \\ 0 & 0 & 0 & 1 & 0 & 0 & 0 & 0 \\ 0 & 0 & 0 & 0 & 1 & 0 & 0 & 0 \\ 0 & 0 & 0 & 0 & 0 & 1 & 0 & 0 \\ 0 & 0 & 0 & 0 & 0 & 0 & 0 & \mathbf{1} \\ 0 & 0 & 0 & 0 & 0 & 0 & \mathbf{1} & 0 \end{bmatrix} \tag{3}
$$

Deutsch [3] extended the work of Toffoli to the quantum domain and introduced a quantum universal gate on three qubits. The core of this gate is specified by the matrix

$$
\boldsymbol{\Delta}(\alpha) = \begin{bmatrix} i\cos(\alpha) & \sin(\alpha) \\ \sin(\alpha) & i\cos(\alpha) \end{bmatrix}, \tag{4}
$$

leading to the parametric Deutsch universal gate, which in the matrix representation is shown in (5), where $\boldsymbol{I}_2$ denotes the 2×2 identity matrix and $\boldsymbol{0}_2$ is a 2×2 matrix with all its entries equal to 0. Note that if $\alpha = \pi/2$, then $\boldsymbol{D}(\pi/2)$ looks like $\boldsymbol{T}$, but in the quantum domain.

$$
\boldsymbol{D}(\alpha) = \begin{bmatrix} \boldsymbol{I}_2 & \boldsymbol{0}_2 & \boldsymbol{0}_2 & \boldsymbol{0}_2 \\ \boldsymbol{0}_2 & \boldsymbol{I}_2 & \boldsymbol{0}_2 & \boldsymbol{0}_2 \\ \boldsymbol{0}_2 & \boldsymbol{0}_2 & \boldsymbol{I}_2 & \boldsymbol{0}_2 \\ \boldsymbol{0}_2 & \boldsymbol{0}_2 & \boldsymbol{0}_2 & \boldsymbol{\Delta}(\alpha) \end{bmatrix}, \tag{5}
$$

The main difference is that in (3), rows and columns are ordered with 3-tuples of Boolean components, meanwhile in (5) rows and columns are ordered according to the Kronecker product of the vectors representing (in a Hilbert space) the *control* and *target* quantum bits.

From (5) it is simple to see that $\boldsymbol{D}$ is a controlled-controlled-$\boldsymbol{\Delta}$ gate, meaning that (1) has its quantum analogue.

The EXOR view however is only valid for the reversible Boolean Toffoli gate, since the closest to (2) in the *p*-valued case is

$$
\text{Not}(target_{in}) = (p-1)(target_{in} + 1) \bmod p. \tag{6}
$$

The $(target_{in} + 1) \bmod p$ could be obtained with a gate, which however, should have to be controlled-controlled by the two control signals, i.e., the original problem appears again in the intended solution. The same applies in the case of a quantum gate for the basis vectors representing eigenstates of a Hilbert space, except that the scaling by $(p-1)$ will be done by a corresponding permutation matrix. (See Eqs. (12), (13)). Since the symbol of the Toffoli gate (Fig. 1) will frequently be also used for quantum-Toffoli gates, The EXOR-view may well be a dangerous source of confusion in a *p*–valued domain. This, however does not contradict the realization of a *p*–valued reversible gate with the functionality

$$
target_{out} = (target_{in} + control_1 \cdot control_2) \bmod p \tag{7}
$$

to work in GF(*p*) [7], except that it is not a Toffoli gate.

For the sake of completeness there are two aspects still to be mentioned in this introduction. Some years after the publication of the pioneering result of Deutsch, an important group of scientists worked together and proved [2] that there are complete sets of quantum gates working on only two qubits. The second aspect to be mentioned is the different language of Switching Theory (binary or multiple-valued), short ST, and Quantum Computing, short, QC. For instance, in ST, variables have scalar Boolean or integer values (according to the domain), meanwhile in QC variables are vectors representing states of a Hilbert space, normally expressed in the Dirac notation [4]. (See e.g. Eqs. (10) and (11)). In ST the cardinality of the set of possible values of the variables represents its valuedness. In QC, depending on the physical environment being considered, is the number of "levels" (or "dimensions") that determines the valuedness of the system [16]. A concept of memory is used in QC, which is totally different to the one in ST. For instance "*We consider the extension of universal quantum logic to the multivalued domain, where the unit of memory is the qudit, a d-dimensional quantum system with the basis states* $|0\rangle$, $|1\rangle$, …, $|d{-}1\rangle$." (Quotation from [16], where the Dirac notation is used to represent the states.).

## II DESIGN OF A QUANTUM *p*-VALUED TOFFOLI GATE

In this section the generalization of a Toffoli gate as a (conjunctive) controlled-controlled-NOT in a *p*-level quantum domain will be discussed, where *p* is a prime larger than 2. In analogy with the binary qubits [20], the basic information units will be called *qupits* [1].

The unitary matrices ***NOT*** for the involutive symmetric negation and the Pauli matrix ***X*** [17] in the *p*-valued domain are

$$\boldsymbol{NOT} = \begin{bmatrix} 0\ 0 \dots & \dots 0\ 1 \\ 0\ 0 \dots & \dots 1\ 0 \\ \dots \dots \dots & \dots \dots \dots \\ \dots \dots \dots & \dots \dots \dots \\ 0\ 1 \dots & \dots 0\ 0 \\ 1\ 0 \dots & \dots 0\ 0 \end{bmatrix} \tag{8}$$

and

$$\boldsymbol{X} = \begin{bmatrix} 0 & 0 & & 0 & 1 \\ 1 & 0 & \cdots & 0 & 0 \\ 0 & 1 & & 0 & 0 \\ \vdots & & \ddots & \vdots & \\ 0 & 0 & \cdots & 1 & 0 \end{bmatrix}, \tag{9}$$

respectively.

Notice that the controlled-***X*** gates, when activated, increase by 1 (mod *p*) the state of their target qupits, as illustrated bellow.

$$\begin{bmatrix} 0 & 0 & & 0 & 1 \\ 1 & 0 & \cdots & 0 & 0 \\ 0 & 1 & & 0 & 0 \\ \vdots & & \ddots & \vdots & \\ 0 & 0 & \cdots & 1 & 0 \end{bmatrix} \cdot \begin{bmatrix} 1 \\ 0 \\ 0 \\ \dots \\ 0 \end{bmatrix} = \begin{bmatrix} 0 \\ 1 \\ 0 \\ \dots \\ 0 \end{bmatrix} \quad \Rightarrow \boldsymbol{X}\cdot|0\rangle = |1\rangle, \tag{10}$$

and

$$\begin{bmatrix} 0 & 0 & & 0 & 1 \\ 1 & 0 & \cdots & 0 & 0 \\ 0 & 1 & & 0 & 0 \\ \vdots & & \ddots & \vdots & \\ 0 & 0 & \cdots & 1 & 0 \end{bmatrix} \cdot \begin{bmatrix} 0 \\ 0 \\ 0 \\ \dots \\ 1 \end{bmatrix} = \begin{bmatrix} 1 \\ 0 \\ 0 \\ \dots \\ 0 \end{bmatrix} \quad \Rightarrow \boldsymbol{X}\cdot|p{-}1\rangle = |0\rangle. \tag{11}$$

respectively

Furthermore let $\overleftarrow{\boldsymbol{P}}$ denote the matrix to scale a vector state by ($p$ –1). For instance,

$$\overleftarrow{\boldsymbol{P}} \cdot |1\rangle = \begin{bmatrix} 1 & 0 & & 0 & 0 \\ 0 & 0 & \cdots & 0 & 1 \\ 0 & 0 & & 1 & 0 \\ \vdots & & \ddots & & \vdots \\ 0 & 1 & \cdots & 0 & 0 \end{bmatrix} \cdot \begin{bmatrix} 0 \\ 1 \\ 0 \\ \dots \\ 0 \end{bmatrix} = \begin{bmatrix} 0 \\ 0 \\ 0 \\ \dots \\ 1 \end{bmatrix} = |p-1\rangle, \tag{12}$$

and

$$\cdot \boldsymbol{X} = \begin{bmatrix} 1 & 0 & & 0 & 0 \\ 0 & 0 & \cdots & 0 & 1 \\ 0 & 0 & & 1 & 0 \\ \vdots & & \ddots & & \vdots \\ 0 & 1 & \cdots & 0 & 0 \end{bmatrix} \cdot \begin{bmatrix} 0 & 0 & & 0 & 1 \\ 1 & 0 & \cdots & 0 & 0 \\ 0 & 1 & & 0 & 0 \\ \vdots & & \ddots & & \vdots \\ 0 & 0 & \cdots & 1 & 0 \end{bmatrix} = $$
$$= \begin{bmatrix} 0 & 0 & & 0 & 1 \\ 0 & 0 & \cdots & 1 & 0 \\ 0 & 0 & & 0 & 0 \\ \vdots & & \ddots & & \vdots \\ 1 & 0 & \cdots & 0 & 0 \end{bmatrix} = \boldsymbol{NOT}. \tag{13}$$

A realization of the $p$-valued Toffoli gate following a suggestion of [1], extensively discussed in [13] is shown in Fig. 2, where $\boldsymbol{N}$ represents the $\boldsymbol{NOT}$ matrix. All elementary quantum gates are Muthukrishnan-Stroud gates [16], which are activated if the control qupit is in state $|\, p–1\rangle$ otherwise being inhibited and behaving as the identity.

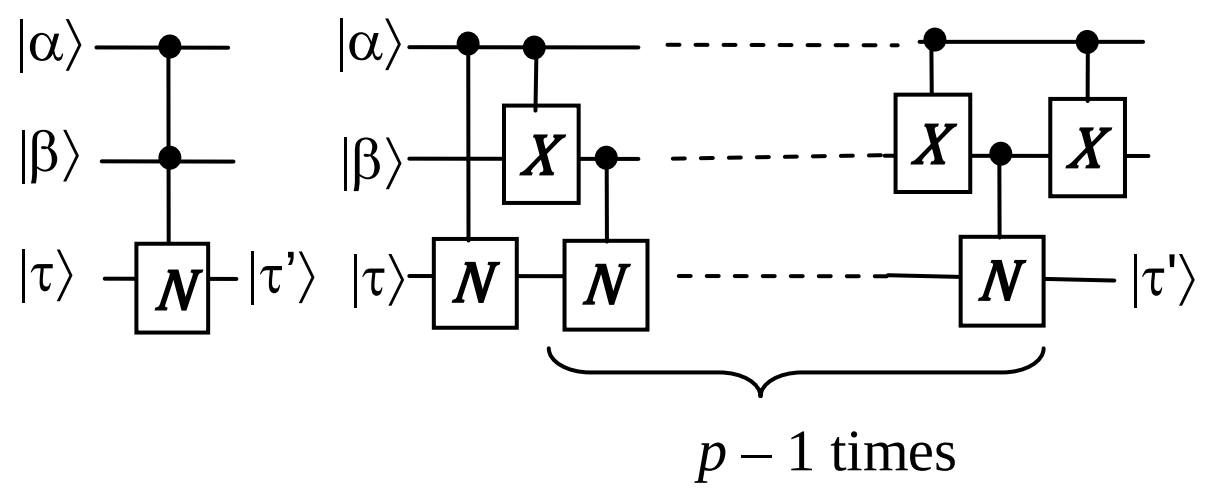

Fig. 2 Quantum realization of a $p$-valued Toffoli gate controlled by $|\alpha\rangle$ ***and*** $|\beta\rangle$.

Analysis of behavior:

Let $\Gamma = (\{|p-1\rangle\}, \Omega\ \}$ denote a partition of the set of basic states, where $\Omega = \{|0\rangle, |1\rangle, ..., |p-2\rangle\}$

- If both $|\alpha\rangle$ and $|\beta\rangle$ are in $\Omega$, all elementary MS gates are inhibited and therefore the Toffoli gate behaves as the identity.
- If $|\alpha\rangle = |p-1\rangle$ and $|\beta\rangle$ is in $\Omega$, the first controlled-$\boldsymbol{N}$ and all controlled-$\boldsymbol{X}$ MS gates are activated. Let $0 < k < p$. If $|\beta\rangle = |p-1-k\rangle$, after the $k$-th controlled $\boldsymbol{X}$, it will reach the state $|p-1\rangle$ and otherwise it will remain in $\Omega$. The corresponding controlled $\boldsymbol{N}$ gate will be activated. This activated controlled-$\boldsymbol{N}$ gate together with the one activated by $|\alpha\rangle$ lead to $\boldsymbol{N}^2 = \boldsymbol{I}$. The Toffoli gate behaves as the identity. Notice that besides the $p$ – 1 controlled-$\boldsymbol{X}$ MS gates that drive the controlled-$\boldsymbol{N}$ gates, there is one additional $\boldsymbol{X}$ gate controlled by $|\alpha\rangle$. This means that from the input to the output, $|\beta\rangle$ will be shifted by "$p$ mod $p$", i.e. the state of $|\beta\rangle$ will be restored.
- If $|\alpha\rangle$ is in $\Omega$ and $|\beta\rangle = |p-1\rangle$, all controlled-$\boldsymbol{X}$ gates are inhibited and all $p$ – 1 controlled-$\boldsymbol{N}$ gates will be activated. Since $p$ – 1 is even and $\boldsymbol{N}$ is its own inverse, then $\boldsymbol{N}^{p-1} = \boldsymbol{I}$. The Toffoli gate behaves as the identity.

- If $|\alpha\rangle = |p-1\rangle$ and $|\beta\rangle = |p-1\rangle$, the first controlled-***N*** and all controlled-***X*** MS gates are activated. $|\beta\rangle$ will be shifted first to $|0\rangle$, then to $|1\rangle$, and further step by step until reaching $|p-2\rangle$. All corresponding controlled-***N*** gates will be inhibited. Only the ***N*** gate controlled by $|\alpha\rangle$ will be active. The Toffoli gate behaves as the intended negation.

The quantum realization of a Toffoli gate with disjunctive control, means that if $|\alpha\rangle$ ***or*** $|\beta\rangle$ are in the state $|p-1\rangle$, then the Toffoli gate behaves as a negation, otherwise as the identity, is shown in Fig. 3, where black triangles are used instead of dots [11] to distinguish this gate (if needed) from the one with conjunctive control. The realization is very similar to that shown in Fig. 2, except that the first controlled-***N*** driven by $|\alpha\rangle$ is moved –(its place is shown with dashed lines in Fig. 3)– and placed as an additional controlled-***N*** driven by the restored $|\beta\rangle$ (shown shaded in Fig. 3). Under these conditions the Toffoli gate consists of *p* hierarchical pairs of controlled-***X*** and controlled-***N*** MS gates.

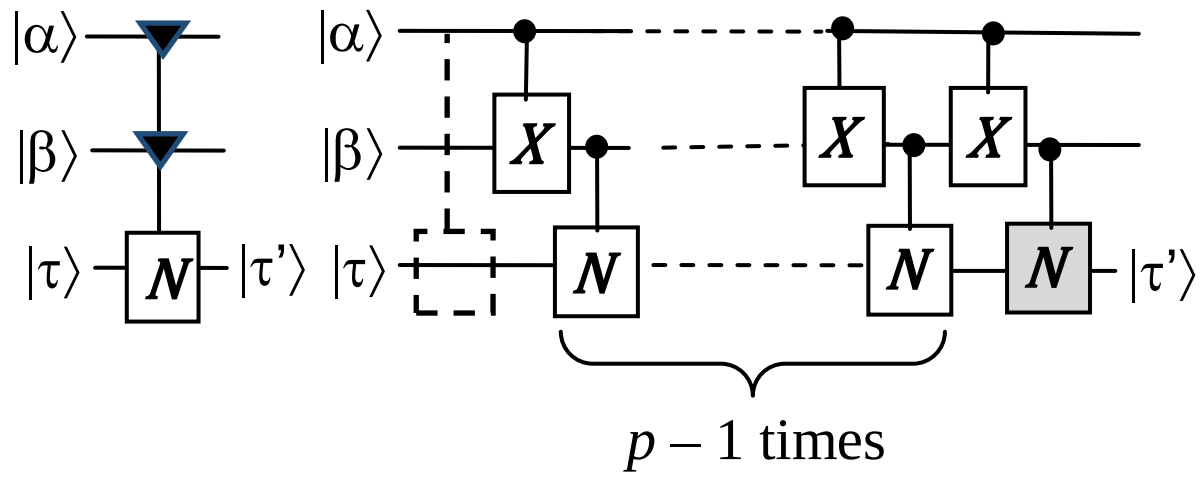

Fig. 3 Quantum realization of a *p*-valued Toffoli gate controlled by $|\alpha\rangle$ ***or*** $|\beta\rangle$.

Analysis of behavior:

- If both $|\alpha\rangle$ and $|\beta\rangle$ are in $\Omega$ all elementary MS gates are inhibited and therefore the Toffoli gate behaves as the identity.
- If $|\alpha\rangle$ is in $\Omega$ and $|\beta\rangle = |p-1\rangle$, all controlled-***X*** are inhibited (behaving as identity), and all *p* controlled-***N*** gates are active, leading to $\boldsymbol{N}^p = \boldsymbol{N}$, since *p* is odd and ***N*** is selfinverse.
- If $|\alpha\rangle = |p-1\rangle$ and $|\beta\rangle$ is in $\Omega$, all controlled-***X*** gates are active and will shift $|\beta\rangle$ over all states of $\Omega \cup \{|p-1\rangle\}$, reaching at only one place the state $|p-1\rangle$. The corresponding controlled-***N*** gate will be activated. Similarly if $|\beta\rangle = |p-1\rangle$, the (new) last controlled-***N*** gate will be active. In both cases the behaviour of the Toffoli gate will be that of a (controlled-controlled)-negation.

## III Design of a quantum *p*-valued Deutsch gate

Deutsch introduced in [3] a large family of universal conjunctive controlled-controlled-***Q*** quantum gates, with ***Q*** specified by a unitary matrix. (***Q*** is unitary if $\boldsymbol{Q} \cdot \boldsymbol{Q}^{\dagger} = \boldsymbol{I}$, where $\boldsymbol{Q}^{\dagger}$ denotes the adjoint (transposed and complex conjugate) of ***Q***. It becomes simple to understand that $\boldsymbol{Q}^{\dagger} = \boldsymbol{Q}^{-1}$).

A generalization of the Deutsch quantum gate to a *p*-valued domain was shortly introduced in [1] and thoroughly analyzed in [13]. The structure of the gate is shown in Fig. 4, where it may be seen that $2p + 1$ elementary MS gates are needed.

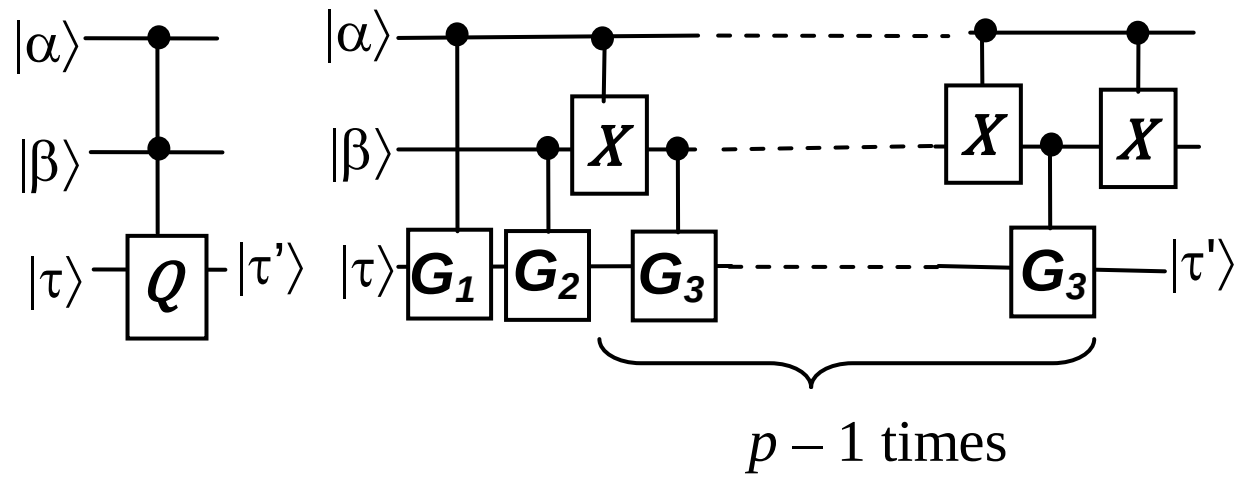

Fig. 4 Scheme for the quantum realization of a $p$-valued Deutsch gate.

The specification of the auxiliary gates $\boldsymbol{G}_1$, $\boldsymbol{G}_2$ and $\boldsymbol{G}_3$ both for the conjunctive (|α⟩ ***and*** |β⟩) control as well as for the disjunctive (|α⟩ ***or*** |β⟩) control was deduced in [13] and is summarized in Table 1.

| Table 1. Auxiliary gates for the quantum realization of the $p$-valued Deutsch gate | | | |
|---|---|---|---|
| Control | $\boldsymbol{G}_1$ | $\boldsymbol{G}_2$ | $\boldsymbol{G}_3$ |
| \|α⟩ ***and*** \|β⟩ | $\sqrt[p]{\boldsymbol{Q}}$ | $(\sqrt[p]{\boldsymbol{Q}})^{p-1}$ | $(\sqrt[p]{\boldsymbol{Q}})^{-1}$ |
| \|α⟩ ***or*** \|β⟩ | $(\sqrt[p]{\boldsymbol{Q}})^{p-1}$ | $\sqrt[p]{\boldsymbol{Q}}$ | $\sqrt[p]{\boldsymbol{Q}}$ |

In [23] it was shown that any non-singular matrix has a $p$-th root. From the mathematical point of view, the calculation of the $p$-th root of a matrix will be based on Schur's triangularization theorem [5], by which any nonsingular matrix is similar to an upper triangular matrix.

$$\boldsymbol{Q} = \boldsymbol{S}\cdot\boldsymbol{U}\cdot\boldsymbol{S}^{-1}, \tag{14}$$

where $\boldsymbol{S}$ is unitary and $\boldsymbol{U}$ is an upper triangular matrix. Moreover if $f(\boldsymbol{Q})$ is defined, then $f(\boldsymbol{Q}) = \boldsymbol{S}\cdot f(\boldsymbol{U})\cdot\boldsymbol{S}^{-1}$ [19], from where $\sqrt[p]{\boldsymbol{Q}} = \boldsymbol{S}\cdot\sqrt[p]{\boldsymbol{U}}\cdot\boldsymbol{S}^{-1}$. Furthermore the $p$-th root of a triangular matrix may be iteratively calculated [23]. From the practical point of view, the $p$-th root of a non-singular matrix may be obtained in Matlab or Scilab by using the simple statement `pth_root_of_Q = expm[(1/p)*logm(Q)]`. In [13] it was shown that the $p$-th root of $\boldsymbol{Q}$ is unitary, and since unitary matrices form a multiplicative group[1], then its inverse as well as an integer power of it are also unitary. The unitary character of matrices is a necessary condition for quantum realizability [9] considering that unitary operators preserve the probabilities for quantum transitions [6].

Analysis of behavior of the Deutsch gate under conjunctive control (|α⟩ ***and*** |β⟩):

- If both |α⟩ and |β⟩ are in Ω, all elementary MS gates are inhibited and therefore the Deutsch gate behaves as the identity.

[1] If $\boldsymbol{A}$ and $\boldsymbol{B}$ are $n\times n$ unitary matrices, $\boldsymbol{A}\cdot\boldsymbol{A}^\dagger = \boldsymbol{I} \Rightarrow \boldsymbol{A}^\dagger = \boldsymbol{A}^{-1}$. From $\boldsymbol{A}\cdot\boldsymbol{A}^{-1} = \boldsymbol{I} \Rightarrow (\boldsymbol{A}\cdot\boldsymbol{A}^{-1})^\dagger = \boldsymbol{A}^\dagger\cdot(\boldsymbol{A}^{-1})^\dagger = \boldsymbol{A}^{-1}\cdot(\boldsymbol{A}^{-1})^\dagger = \boldsymbol{I}$, i.e. the inverse of a unitary matrix is unitary. $(\boldsymbol{A}\cdot\boldsymbol{B})^\dagger\cdot(\boldsymbol{A}\cdot\boldsymbol{B}) = \boldsymbol{B}^\dagger\cdot\boldsymbol{A}^\dagger\cdot\boldsymbol{A}\cdot\boldsymbol{B} = \boldsymbol{B}^\dagger\cdot\boldsymbol{B} = \boldsymbol{I}$, i.e. the product of unitary matrices is unitary. Finally, the product of matrices is associative. This characterizes a group.

- If $|\alpha\rangle = |p{-}1\rangle$ and $|\beta\rangle$ is in $\Omega$, $\boldsymbol{G}_1$ and all controlled-$\boldsymbol{X}$ gates are active. This will shift $|\beta\rangle$ over all states of $\Omega \cup \{|p{-}1\rangle\}$, reaching at only one place the state $|p{-}1\rangle$. The corresponding controlled-$\boldsymbol{G}_3$ gate will be activated. The Deutsch gate will behave as $\boldsymbol{G}_1 \cdot \boldsymbol{G}_3 = \sqrt[p]{\boldsymbol{Q}} \cdot \left(\sqrt[p]{\boldsymbol{Q}}\right)^{-1} = \boldsymbol{I}$.
- If $|\alpha\rangle$ is in $\Omega$ and $|\beta\rangle = |p{-}1\rangle$, all controlled-$\boldsymbol{X}$ are inhibited (behaving as identity). $\boldsymbol{G}_2$ and all $p$ controlled-$\boldsymbol{G}_3$ gates are activated, leading to $\boldsymbol{G}_2 \cdot (\boldsymbol{G}_3)^{p-1} = \left(\sqrt[p]{\boldsymbol{Q}}\right)^{p-1} \left(\left(\sqrt[p]{\boldsymbol{Q}}\right)^{-1}\right)^{p-1} = \boldsymbol{I}$.
- If $|\alpha\rangle = |\beta\rangle = |p{-}1\rangle$, all controlled-$\boldsymbol{X}$ gates are activated, but $|\beta\rangle$ will be shifted up to $|p{-}2\rangle$, i.e., all $\boldsymbol{G}_3$ gates will be inhibited. Only $\boldsymbol{G}_1$ and $\boldsymbol{G}_2$ will be activated leading to the global behaviour $\boldsymbol{G}_1 \cdot \boldsymbol{G}_2 = \sqrt[p]{\boldsymbol{Q}} \cdot \left(\sqrt[p]{\boldsymbol{Q}}\right)^{p-1} = \left(\sqrt[p]{\boldsymbol{Q}}\right)^{p} = \mathbf{Q}$, as desired.

Analysis of behavior of the Deutsch gate under disjunctive control ($|\alpha\rangle$ ***or*** $|\beta\rangle$):

- If both $|\alpha\rangle$ and $|\beta\rangle$ are in $\Omega$, all elementary MS gates are inhibited and therefore the Deutsch gate behaves as the identity.
- If $|\alpha\rangle$ is in $\Omega$ and $|\beta\rangle = |p{-}1\rangle$, all controlled-$\boldsymbol{X}$ are inhibited (behaving as identity). $\boldsymbol{G}_2$ and all $p{-}1$ controlled-$\boldsymbol{G}_3$ gates are activated, leading to $\boldsymbol{G}_2 \cdot (\boldsymbol{G}_3)^{p-1} = \sqrt[p]{\boldsymbol{Q}} \cdot \left(\sqrt[p]{\boldsymbol{Q}}\right)^{p-1} = \left(\sqrt[p]{\boldsymbol{Q}}\right)^{p} = \boldsymbol{Q}$.
- If $|\alpha\rangle = |p{-}1\rangle$ and $|\beta\rangle$ is in $\Omega$, $\boldsymbol{G}_1$ and all controlled-$\boldsymbol{X}$ gates are active and will shift $|\beta\rangle$ over all states of $\Omega \cup \{|p{-}1\rangle\}$, reaching at only one place the state $|p{-}1\rangle$. The corresponding controlled-$\boldsymbol{G}_3$ gate will be activated leading to the global behaviour $\boldsymbol{G}_1 \cdot \boldsymbol{G}_3 = \left(\sqrt[p]{\boldsymbol{Q}}\right)^{p-1} \cdot \left(\sqrt[p]{\boldsymbol{Q}}\right) = \left(\sqrt[p]{\boldsymbol{Q}}\right)^{p} = \boldsymbol{Q}$.
- Similarly if $|\alpha\rangle = |p{-}1\rangle$ and $|\beta\rangle = |p{-}1\rangle$, $\boldsymbol{G}_1$, $\boldsymbol{G}_2$ and all controlled-$\boldsymbol{X}$ gates will be activated. However, the controlled-$\boldsymbol{X}$ gates will shift $|\beta\rangle$ through $\Omega$, hence all $\boldsymbol{G}_3$ gates will remain inhibited. The global behaviour will be $\boldsymbol{G}_1 \cdot \boldsymbol{G}_2 = \left(\sqrt[p]{\boldsymbol{Q}}\right)^{p-1} \cdot \left(\sqrt[p]{\boldsymbol{Q}}\right) = \left(\sqrt[p]{\boldsymbol{Q}}\right)^{p} = \boldsymbol{Q}$.

The analysis shows that it is enough that one of the control qupits is in state $|p{-}1\rangle$ for the Deutsch gate to become active.

## IV *p*-VALUED DEUTSCH GATES WITH MIXED POLARITY

The concept of mixed polarities for reversible circuits, borrowed from Reed Muller expressions [18], was possibly first introduced in [21]. The concept is used to express that in the case of (binary) quantum Deutsch gates with two control qubits, the gate is active when any one of them is in *state* $|1\rangle$ and the other in state $|0\rangle$. (Similarly for reversible Toffoli gates, where one control signal has the *value* 1 and the other the value 0. See e.g. [8], [10], and for more than two control signals, see e.g. [21], [12].)

To extend the concept of mixed polarity to an MS-based *p*-valued quantum Deutsch gate has the particular difficulty that one control qupit should be in the state $|p\text{-}1\rangle$ and the other one, in *any* state of $\Omega$ to activate the gate. In the binary case, in the symbolic representation of the gate it has been adopted the convention of using "black dots" to indicate that the corresponding the control qubit should be in state $|1\rangle$ to (contribute to) activate the controlled gate. Similarly, "white dots" are used to indicate that the control qubit should be in state $|0\rangle$ to (contribute to) activate the gate. For a gate in a *p*–levels system a similar convention will be used, except that instead of "white dots", "white diamonds" will be used, meaning that any state other than $|p–1\rangle$ will be effective. Fig. 5 shows the symbols of the two cases that will be analyzed below.

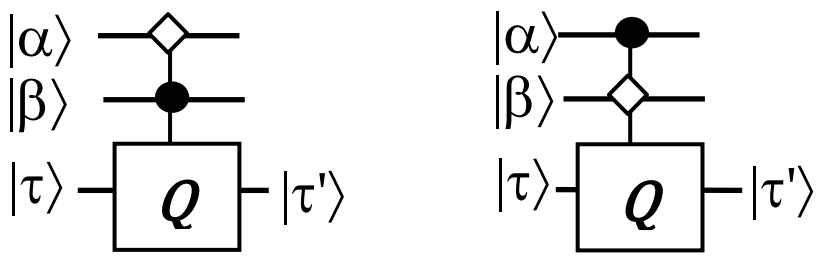

Fig. 5 Quantum $p$–valued Deutsch gates with conjunctive control under mixed polarity.
Left: Gate active when $|\alpha\rangle \in \Omega$ and $|\beta\rangle = |p–1\rangle$.
Right: Gate active when $|\alpha\rangle = |p–1\rangle$ and $|\beta\rangle \in \Omega$.

The analysis of behavior of the Deutsch gate shown in Fig. 4 under conjunctive control with mixed polarity follows the same steps as in the former cases and is summarized in Table 2. (In Table 2, $\{\boldsymbol{X}\}$ and $\{\boldsymbol{G}_3\}$ denote all the $\boldsymbol{X}$ and $\boldsymbol{G}_3$ gates, respectively.)

| Table 2: Behavior of the Deutsch gate under conjunctive control with different polarities | | | | |
|---|---|---|---|---|
| $|\alpha\rangle$<br>$|\beta\rangle$ | $|p-1\rangle$<br>$\in \Omega$ | $\in \Omega$<br>$|p-1\rangle$ | $|p-1\rangle$<br>$|p-1\rangle$ | $\in \Omega$<br>$\in \Omega$ |
| $\{\boldsymbol{X}\}$ | $\{\boldsymbol{X}\}$ | $\{\boldsymbol{I}\}$ | $\{\boldsymbol{X}\}$ | $\{\boldsymbol{I}\}$ |
| $\boldsymbol{G}_1$ | $\boldsymbol{G}_1$ | $\boldsymbol{I}$ | $\boldsymbol{G}_1$ | $\boldsymbol{I}$ |
| $\boldsymbol{G}_2$ | $\boldsymbol{I}$ | $\boldsymbol{G}_2$ | $\boldsymbol{G}_2$ | $\boldsymbol{I}$ |
| $\{\boldsymbol{G}_3\}$ | $\boldsymbol{G}_3$ | $(\boldsymbol{G}_3)^{p-1}$ | $\{\boldsymbol{I}\}$ | $\{\boldsymbol{I}\}$ |
| 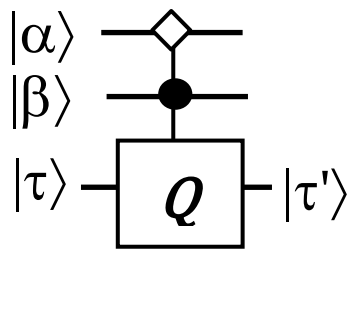 | $\boldsymbol{G}_1\boldsymbol{G}_3 = \boldsymbol{I}$ | $\boldsymbol{G}_2(\boldsymbol{G}_3)^{p-1} = \boldsymbol{Q}$ | $\boldsymbol{G}_1\boldsymbol{G}_2 = \boldsymbol{I}$ | $\{\boldsymbol{I}\}$ |
| 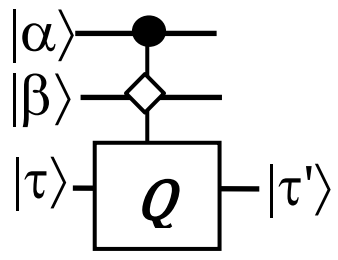 | $\boldsymbol{G}_1\boldsymbol{G}_3 = \boldsymbol{Q}$ | $\boldsymbol{G}_2(\boldsymbol{G}_3)^{p-1} = \boldsymbol{I}$ | $\boldsymbol{G}_1\boldsymbol{G}_2 = \boldsymbol{I}$ | $\{\boldsymbol{I}\}$ |

From Table 2, for the "white-black" Deutsch gate holds that $\boldsymbol{G}_1\boldsymbol{G}_3 = \boldsymbol{G}_1\boldsymbol{G}_2 = \boldsymbol{I}$, from where $\boldsymbol{G}_2 = \boldsymbol{G}_3$ and $\boldsymbol{G}_1 = (\boldsymbol{G}_3)^{-1}$. Introducing these equalities in the equation $\boldsymbol{G}_2(\boldsymbol{G}_3)^{p-1} = \boldsymbol{Q}$ follows that

$$\boldsymbol{G}_2 = \boldsymbol{G}_3 = \sqrt[p]{\boldsymbol{Q}}. \tag{15}$$

and

$$\boldsymbol{G}_1 = \sqrt[p]{\boldsymbol{Q}^{-1}}. \tag{16}$$

As mentioned earlier, these are unitary matrices.

For the "black-white" Deutsch gate holds that $\boldsymbol{G}_1\boldsymbol{G}_2 = \boldsymbol{I}$, from where $\boldsymbol{G}_2 = (\boldsymbol{G}_1)^{-1}$. Similarly, from $\boldsymbol{G}_2(\boldsymbol{G}_3)^{p-1} = \boldsymbol{I}$ follows that $(\boldsymbol{G}_2)^{-1} = (\boldsymbol{G}_3)^{p-1}$ and $\boldsymbol{G}_1 = (\boldsymbol{G}_3)^{p-1}$. Introducing these equalities in the equation $\boldsymbol{G}_1\boldsymbol{G}_3 = \boldsymbol{Q}$ leads to

$$(\boldsymbol{G}_3)^{p-1}\,\boldsymbol{G}_3 = (\boldsymbol{G}_3)^{p} = \boldsymbol{Q}, \tag{17}$$

from where

$$\boldsymbol{G}_3 = \sqrt[p]{\boldsymbol{Q}}\,, \tag{18}$$

$$\boldsymbol{G}_1 = (\boldsymbol{G}_3)^{p-1} = \left(\sqrt[p]{\boldsymbol{Q}}\;\right)^{p-1}, \tag{19}$$

$$\boldsymbol{G}_2 = (\boldsymbol{G}_1)^{-1} = \left(\sqrt[p]{\boldsymbol{Q}^{-1}}\right)^{p-1}. \tag{20}$$

The three matrices are unitary.

Since whenever both $|\alpha\rangle$ and $|\beta\rangle$ are in $\Omega$ all the component MS gates and, therefore, the Deutsch gate behaves as the identity, an unexpected trivialization occurs if a disjunctive control under mixed polarity is intended.

Consider first the case that the Deutsch gate (of Fig. 4) should be active when $|\alpha\rangle = |p-1\rangle$ ***or*** $|\beta\rangle \in \Omega$. The activating control pairs are ($|\alpha\rangle = |p-1\rangle$, $|\beta\rangle \in \Omega$) requiring $\boldsymbol{G}_1\boldsymbol{G}_3 = \boldsymbol{Q}$, and ($|\alpha\rangle = |p-1\rangle$, $|\beta\rangle = |p-1\rangle$) leading to $\boldsymbol{G}_1\boldsymbol{G}_2 = \boldsymbol{Q}$. The inhibiting control pair is ($|\alpha\rangle \in \Omega$, $|\beta\rangle = |p-1\rangle$). Since in this case all controlled-$\boldsymbol{X}$ are inhibited, but $\boldsymbol{G}_2$ and all $p-1$ controlled-$\boldsymbol{G}_3$ gates are activated, the equality $\boldsymbol{G}_2(\boldsymbol{G}_3)^{p-1} = \boldsymbol{I}$ must be satisfied.

$$\boldsymbol{G}_1\boldsymbol{G}_2 = \boldsymbol{G}_1\boldsymbol{G}_3 = \boldsymbol{Q} \Rightarrow \boldsymbol{G}_2 = \boldsymbol{G}_3 . \tag{21}$$

Introducing (21) into $\boldsymbol{G}_2(\boldsymbol{G}_3)^{p-1} = \boldsymbol{I}$ one obtains

$$(\boldsymbol{G}_3)^p = \boldsymbol{I}, \tag{22}$$

which leads to $\boldsymbol{G}_2 = \boldsymbol{G}_3 = \boldsymbol{I}$ and hence $\boldsymbol{G}_1 = \boldsymbol{Q}$. This means that the whole Deutsch gate is reduced to a single simple MS gate controlled by $|\alpha\rangle$, *independently* of the state of $|\beta\rangle$.

Analogously, if the gate should be active when $|\alpha\rangle \in \Omega$ ***or*** $|\beta\rangle = |p-1\rangle$, the Deutsch gate will be reduced to a single MS gate realizing $\boldsymbol{Q}$ controlled by $|\beta\rangle$, *independently* of the state of $|\alpha\rangle$.

It is possible, however, to have an "exclusive-disjunct" control with mixed polarity, this meaning that the Deutsch gate would active when [$|\alpha\rangle = |p-1\rangle$ ***and*** $|\beta\rangle \in \Omega$] ***or*** [$|\alpha\rangle \in \Omega$ ***and*** $|\beta\rangle = |p-1\rangle$], i.e. when $|\alpha\rangle$ and $|\beta\rangle$ are in states from different blocks of the partition $\Gamma$. The gate should be inhibited, when both $|\alpha\rangle$ and $|\beta\rangle = |p-1\rangle$. Formally, the activation conditions are given by

$$\boldsymbol{G}_1\boldsymbol{G}_3 = \boldsymbol{Q}, \tag{23}$$

and

$$\boldsymbol{G}_2(\boldsymbol{G}_3)^{p-1} = \boldsymbol{Q}. \tag{24}$$

The inhibiting condition is

$$\boldsymbol{G}_1\boldsymbol{G}_2 = \boldsymbol{I}, \tag{25}$$

From (23) and (25) follows that

$$(\boldsymbol{G}_2)^{-1}\boldsymbol{G}_3 = \boldsymbol{Q}, \tag{26}$$

and

$$\boldsymbol{G}_3 = \boldsymbol{G}_2\boldsymbol{Q}, \tag{27}$$

which with (24) gives

$$\boldsymbol{G}_2(\boldsymbol{G}_2\boldsymbol{Q})^{p-1} = \boldsymbol{Q}, \tag{28}$$

$$(\boldsymbol{G}_2)^p\boldsymbol{Q}^{p-1} = \boldsymbol{Q}, \tag{29}$$

$$\boldsymbol{G}_2 = \sqrt[p]{\boldsymbol{Q}^{2-p}} . \tag{30}$$

With (25)

$$\boldsymbol{G}_1 = \sqrt[p]{\boldsymbol{Q}^{p-2}}. \tag{31}$$

Finally, with (27),

$$\boldsymbol{G}_3 = \cdot \sqrt[p]{\boldsymbol{Q}^{2-p}} = \sqrt[p]{\boldsymbol{Q}^{2}} . \tag{32}$$

As stated earlier, all involved matrices are unitary.

## V  DESIGN OF A *p*-VALUED DEUTSCH QUANTUM GATE WITH THREE CONTROL QUPITS

For the design of a *p*-valued Deutsch quantum gate with three control qupits, a combination of the schemes discussed in the former sections, with the structures disclosed by Barenco et al. [2] for the binary case will be considered. Fig. 6 shows a first conjecture on the realization of the gate, where A, B, …, G are labels for gates that will be specified by the unitary matrices ***A***, ***B***,…, ***G***. Gates labeled X will be specified by the ***X*** Pauli matrices.

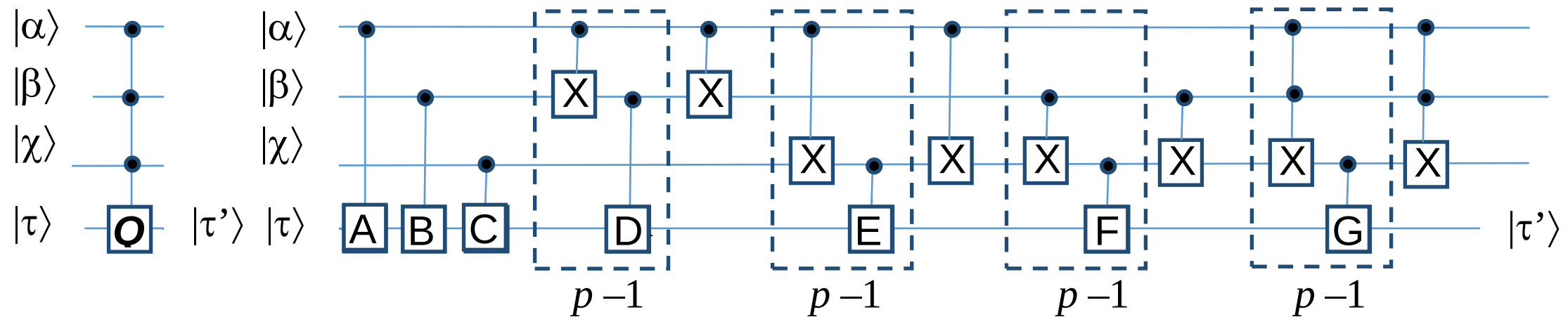

Fig. 6: General scheme conjectured for a Deutsch gate with three controls

Comparing with structures developed in [2] it seems unusual to use controlled-controlled-***X*** gates in the G-block instead of a cascade of two simpler controlled-***X*** gates. The following example, illustrated in Fig. 7 for $p = 5$ shows the difficulties to control the block G with cascades of two ***CX*** gates as compared with the control with ***CCX*** gates. This is summarized in Table 3, where $|\omega\rangle \in \Omega$ .

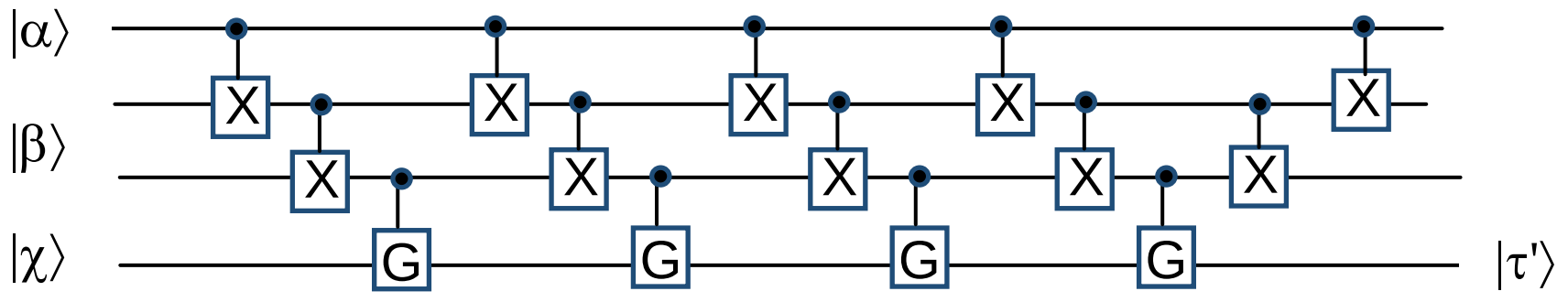

Fig. 7: Block control with cascaded ***CX*** gates when $p = 5$

Table 3: Transfer function of a block under cascaded ***CX*** control compared to ***CCX*** control ($p = 5$)

| Control | | | 1st step | | | 2nd step | | | 3rd step | | | 4th step | | | Transfer | |
|---|---|---|---|---|---|---|---|---|---|---|---|---|---|---|---|---|
| \|α⟩ | \|β⟩ | \|χ⟩ | \|α⟩ | \|β⟩ | \|χ⟩ | \|α⟩ | \|β⟩ | \|χ⟩ | \|α⟩ | \|β⟩ | \|χ⟩ | \|α⟩ | \|β⟩ | \|χ⟩ | 2 ***CX*** | ***CCX*** |
| \|ω⟩ | \|4⟩ | \|0⟩ | \|ω⟩ | \|4⟩ | \|1⟩ | \|ω⟩ | \|4⟩ | \|2⟩ | \|ω⟩ | \|4⟩ | \|3⟩ | \|ω⟩ | \|4⟩ | \|4⟩ | ***G*** | ***I*** |
| | | \|3⟩ | | | \|4⟩ | | | \|0⟩ | | | \|1⟩ | | | \|2⟩ | ***G*** | ***I*** |
| | | \|4⟩ | | | \|0⟩ | | | \|1⟩ | | | \|2⟩ | | | \|3⟩ | ***I*** | ***I*** |
| \|4⟩ | \|0⟩ | \|0⟩ | \|4⟩ | \|1⟩ | \|0⟩ | \|4⟩ | \|2⟩ | \|0⟩ | \|4⟩ | \|3⟩ | \|0⟩ | \|4⟩ | \|4⟩ | \|1⟩ | ***I*** | ***I*** |
| | | \|3⟩ | | | \|3⟩ | | | \|3⟩ | | | \|3⟩ | | | \|4⟩ | ***G*** | ***I*** |
| | | \|4⟩ | | | \|4⟩ | | | \|4⟩ | | | \|4⟩ | | | \|0⟩ | $G^3$ | ***I*** |
| \|4⟩ | \|2⟩ | \|3⟩ | \|4⟩ | \|3⟩ | \|3⟩ | \|4⟩ | \|4⟩ | \|4⟩ | \|4⟩ | \|0⟩ | \|4⟩ | \|4⟩ | \|1⟩ | \|4⟩ | $G^3$ | ***I*** |
| \|4⟩ | \|1⟩ | \|3⟩ | \|4⟩ | \|2⟩ | \|3⟩ | \|4⟩ | \|3⟩ | \|3⟩ | \|4⟩ | \|4⟩ | \|4⟩ | \|4⟩ | \|0⟩ | \|4⟩ | $G^2$ | ***I*** |
| \|4⟩ | \|1⟩ | \|4⟩ | \|4⟩ | \|2⟩ | \|4⟩ | \|4⟩ | \|3⟩ | \|4⟩ | \|4⟩ | \|4⟩ | \|0⟩ | \|4⟩ | \|1⟩ | \|0⟩ | $G^2$ | ***I*** |
| \|4⟩ | \|4⟩ | \|0⟩ | \|4⟩ | \|0⟩ | \|0⟩ | \|4⟩ | \|1⟩ | \|0⟩ | \|4⟩ | \|2⟩ | \|0⟩ | \|4⟩ | \|3⟩ | \|0⟩ | ***I*** | ***G*** |
| | | \|3⟩ | | | \|3⟩ | | | \|3⟩ | | | \|3⟩ | | | \|3⟩ | ***I*** | ***G*** |
| | | \|4⟩ | | | \|4⟩ | | | \|4⟩ | | | \|4⟩ | | | \|4⟩ | $G^4$ | ***I*** |
| \|ω⟩ | \|ω⟩ | \|4⟩ | \|ω⟩ | \|ω⟩ | \|4⟩ | \|ω⟩ | \|ω⟩ | \|4⟩ | \|ω⟩ | \|ω⟩ | \|4⟩ | \|ω⟩ | \|ω⟩ | \|4⟩ | $G^4$ | $G^4$ |

It is simple to see in Table 3, that with the cascaded control, the block may behave as any of the powers of $\boldsymbol{G}$, meanwhile with the $CC\boldsymbol{X}$ control the block behaves a $\boldsymbol{G}$ iff $|\alpha\rangle = |\beta\rangle = |p-1\rangle$ and $|\chi\rangle = |\omega\rangle$, and it behaves as $\boldsymbol{G}^{p-1}$ iff $|\alpha\rangle = |\beta\rangle = |\omega\rangle$ and $|\chi\rangle = |p-1\rangle$, otherwise behaving as the identity.

The behavior of the proposed Deutsch gate, at an abstract level, is summarized in Table 4, where $|\omega\rangle \in \Omega$ and $q = p - 1$. The last two columns express the intended behavior with conjunctive (***and***) and disjunctive (***or***) control, respectively. In the other columns, the following convention is used: for instance E means that in the corresponding block, only one MS-gate is active (not being important which one), meanwhile $E^q$ indicates that all $p-1$ elementary MS-gates of the block are active. Since after every block the control state is recovered, the blocks may be reordered and, therefore, the product of the unitary matrices specifying the elementary gates is commutative.

Table 4. Abstract level behavior of the Deutsch gate of Fig. 6

| | $\lvert\alpha\rangle$ | $\lvert\beta\rangle$ | $\lvert\chi\rangle$ | *A* | *B* | *C* | *D* | *E* | *F* | *G* | *and* | *or* |
|---|---|---|---|---|---|---|---|---|---|---|---|---|
| 0 | $\lvert\omega\rangle$ | $\lvert\omega\rangle$ | $\lvert\omega\rangle$ | $\boldsymbol{I}$ | $\boldsymbol{I}$ | $\boldsymbol{I}$ | $\boldsymbol{I}^q$ | $\boldsymbol{I}^q$ | $\boldsymbol{I}^q$ | $\boldsymbol{I}^q$ | $\boldsymbol{I}$ | $\boldsymbol{I}$ |
| 1 | $\lvert\omega\rangle$ | $\lvert\omega\rangle$ | $\lvert p{-}1\rangle$ | $\boldsymbol{I}$ | $\boldsymbol{I}$ | $\boldsymbol{C}$ | $\boldsymbol{I}^q$ | $\boldsymbol{E}^q$ | $\boldsymbol{F}^q$ | $\boldsymbol{G}^q$ | $\boldsymbol{I}$ | $\boldsymbol{Q}$ |
| 2 | $\lvert\omega\rangle$ | $\lvert p{-}1\rangle$ | $\lvert\omega\rangle$ | $\boldsymbol{I}$ | $\boldsymbol{B}$ | $\boldsymbol{I}$ | $\boldsymbol{D}^q$ | $\boldsymbol{I}^q$ | $\boldsymbol{F}$ | $\boldsymbol{I}^q$ | $\boldsymbol{I}$ | $\boldsymbol{Q}$ |
| 3 | $\lvert p{-}1\rangle$ | $\lvert\omega\rangle$ | $\lvert\omega\rangle$ | $\boldsymbol{A}$ | $\boldsymbol{I}$ | $\boldsymbol{I}$ | $\boldsymbol{D}$ | $\boldsymbol{E}$ | $\boldsymbol{I}^q$ | $\boldsymbol{I}^q$ | $\boldsymbol{I}$ | $\boldsymbol{Q}$ |
| 4 | $\lvert\omega\rangle$ | $\lvert p{-}1\rangle$ | $\lvert p{-}1\rangle$ | $\boldsymbol{I}$ | $\boldsymbol{B}$ | $\boldsymbol{C}$ | $\boldsymbol{D}^q$ | $\boldsymbol{E}^q$ | $\boldsymbol{I}^q$ | $\boldsymbol{G}^q$ | $\boldsymbol{I}$ | $\boldsymbol{Q}$ |
| 5 | $\lvert p{-}1\rangle$ | $\lvert\omega\rangle$ | $\lvert p{-}1\rangle$ | $\boldsymbol{A}$ | $\boldsymbol{I}$ | $\boldsymbol{C}$ | $\boldsymbol{D}$ | $\boldsymbol{I}^q$ | $\boldsymbol{F}^q$ | $\boldsymbol{G}^q$ | $\boldsymbol{I}$ | $\boldsymbol{Q}$ |
| 6 | $\lvert p{-}1\rangle$ | $\lvert p{-}1\rangle$ | $\lvert\omega\rangle$ | $\boldsymbol{A}$ | $\boldsymbol{B}$ | $\boldsymbol{I}$ | $\boldsymbol{I}^q$ | $\boldsymbol{E}$ | $\boldsymbol{F}$ | $\boldsymbol{G}$ | $\boldsymbol{I}$ | $\boldsymbol{Q}$ |
| 7 | $\lvert p{-}1\rangle$ | $\lvert p{-}1\rangle$ | $\lvert p{-}1\rangle$ | $\boldsymbol{A}$ | $\boldsymbol{B}$ | $\boldsymbol{C}$ | $\boldsymbol{I}^q$ | $\boldsymbol{I}^q$ | $\boldsymbol{I}^q$ | $\boldsymbol{I}^q$ | $\boldsymbol{Q}$ | $\boldsymbol{Q}$ |

i) Deduction of the elementary gates for the conjunctive control.

The gate should be active, $|\tau'\rangle = \boldsymbol{Q}\,|\tau\rangle$, iff $|\alpha\rangle = |\beta\rangle = |\chi\rangle = |p-1\rangle$

From the rows 1, 4 and 5 of the table 4 follows that:

$$\boldsymbol{C} = (\boldsymbol{E}^q\boldsymbol{F}^q\boldsymbol{G}^q)^{-1} = (\boldsymbol{B}\boldsymbol{D}^q\boldsymbol{E}^q\boldsymbol{G}^q)^{-1} = (\boldsymbol{A}\boldsymbol{D}\boldsymbol{F}^q\boldsymbol{G}^q)^{-1}, \tag{33}$$

leading to

$$(\boldsymbol{E}^q\boldsymbol{F}^q\boldsymbol{G}^q) = (\boldsymbol{B}\boldsymbol{D}^q\boldsymbol{E}^q\boldsymbol{G}^q) \Rightarrow \boldsymbol{F}^q = \boldsymbol{B}\boldsymbol{D}^q, \tag{34}$$

and

$$(\boldsymbol{B}\boldsymbol{D}^q\boldsymbol{E}^q\boldsymbol{G}^q) = (\boldsymbol{A}\boldsymbol{D}\boldsymbol{F}^q\boldsymbol{G}^q) \Rightarrow \boldsymbol{B}\boldsymbol{D}^q\boldsymbol{E}^q = \boldsymbol{A}\boldsymbol{D}\boldsymbol{F}^q, \tag{35}$$

with (34)

$$\boldsymbol{B}\boldsymbol{D}^q\boldsymbol{E}^q = \boldsymbol{A}\boldsymbol{D}\boldsymbol{F}^q \Rightarrow \boldsymbol{E}^q = \boldsymbol{A}\boldsymbol{D}\,, \tag{36}$$

Since from row 3

$$\boldsymbol{E}^{-1} = \boldsymbol{A}\boldsymbol{D}, \tag{37}$$

then

$$\boldsymbol{E}^q = \boldsymbol{E}^{-1} \Rightarrow \boldsymbol{E} = \boldsymbol{I}. \tag{38}$$

With (38) in (37)

$$\boldsymbol{A}\boldsymbol{D} = \boldsymbol{I} \Rightarrow \boldsymbol{A} = \boldsymbol{D}^{-1}, \tag{39}$$

and with (39) in (33)

$$\boldsymbol{C} = (\boldsymbol{A}\boldsymbol{D}\boldsymbol{F}^q\boldsymbol{G}^q)^{-1} \Rightarrow \boldsymbol{C} = (\boldsymbol{F}^q\boldsymbol{G}^q)^{-1} \Rightarrow \boldsymbol{F}\boldsymbol{G} = \sqrt[q]{\boldsymbol{C}^{-1}}\,. \tag{40}$$

Rows 5, 6 with (38)

$$\boldsymbol{A}\boldsymbol{C}\boldsymbol{D}\boldsymbol{F}^q\boldsymbol{G}^q = \boldsymbol{A}\boldsymbol{B}\boldsymbol{E}\boldsymbol{F}\boldsymbol{G} \Rightarrow \boldsymbol{C}\boldsymbol{D}\boldsymbol{F}^q\boldsymbol{G}^q = \boldsymbol{B}\boldsymbol{F}\boldsymbol{G},$$

With (34) and (40)

$$\boldsymbol{C}\boldsymbol{D}\boldsymbol{D}^q\boldsymbol{G}^q\boldsymbol{B} = \boldsymbol{B}\boldsymbol{F}\boldsymbol{G} \Rightarrow \boldsymbol{D}^p\boldsymbol{G}^q = \boldsymbol{C}^{-1}\sqrt[q]{\boldsymbol{C}^{-1}} = \sqrt[q]{\boldsymbol{C}^{-p}}\,, \tag{41}$$

Rows 2, 4 with (38)

$$\boldsymbol{B}\boldsymbol{D}^q\boldsymbol{F} = \boldsymbol{B}\boldsymbol{C}\boldsymbol{D}^q\boldsymbol{G}^q \Rightarrow \boldsymbol{F} = \boldsymbol{C}\boldsymbol{G}^q, \tag{42}$$

With (42) and (40)

$$\boldsymbol{F}\boldsymbol{G} = \boldsymbol{C}\boldsymbol{G}^p = \sqrt[q]{\boldsymbol{C}^{-1}} \Rightarrow \boldsymbol{G}^p = \boldsymbol{C}^{-1}\sqrt[q]{\boldsymbol{C}^{-1}} = \sqrt[q]{\boldsymbol{C}^{-p}} \Rightarrow \tag{43}$$

$$\Rightarrow \boldsymbol{G} = \sqrt[q]{\boldsymbol{C}^{-1}}, \tag{44}$$

With (44) and (42) $$\boldsymbol{F} = \boldsymbol{CG}^q = \boldsymbol{C}\left(\sqrt[q]{\boldsymbol{C}^{-1}}\right)^{\boldsymbol{q}} = \boldsymbol{C}\left(\sqrt[q]{\boldsymbol{C}^{-q}}\right) = \boldsymbol{CC}^{-1} \Rightarrow \boldsymbol{F} = \boldsymbol{I}. \quad (45)$$

From row 7 $$\boldsymbol{C} = \boldsymbol{QA}^{-1}\boldsymbol{B}^{-1} \Rightarrow \boldsymbol{AB} = \boldsymbol{QC}^{-1}, \quad (46)$$

From row 6, (38), (45) $$\boldsymbol{AB} = \boldsymbol{G}^{-1}, \quad (47)$$

and with (44) $$\boldsymbol{AB} = \left(\sqrt[q]{\boldsymbol{C}^{-1}}\right)^{-\boldsymbol{1}} = \sqrt[q]{\boldsymbol{C}} = \boldsymbol{QC}^{-1} \Rightarrow \boldsymbol{C}\sqrt[q]{\boldsymbol{C}} = \boldsymbol{Q} \Rightarrow \boldsymbol{C} = \sqrt[p]{\boldsymbol{Q}^q}. \quad (48)$$

With (48) in (44) $$\boldsymbol{G} = \sqrt[q]{\boldsymbol{C}^{-1}} = \sqrt[q]{\left(\sqrt[p]{\boldsymbol{Q}^q}\right)^{-1}} = \sqrt[p]{\left(\sqrt[q]{\boldsymbol{Q}^q}\right)^{-1}} \Rightarrow \boldsymbol{G} = \sqrt[p]{\boldsymbol{Q}^{-1}}. \quad (49)$$

Rows 2, 6 and (38), (39), (45) $$\boldsymbol{BD}^q = \boldsymbol{ABG} = \boldsymbol{D}^{-1}\boldsymbol{BG} \Rightarrow \boldsymbol{D}^p = \boldsymbol{G} \Rightarrow \boldsymbol{D} = \sqrt[p]{\boldsymbol{G}} = \sqrt[p^2]{\boldsymbol{Q}^{-1}}. \quad (50)$$

With (39) in (50) $$\boldsymbol{A} = \boldsymbol{D}^{-1} \Rightarrow \boldsymbol{A} = \sqrt[p^2]{\boldsymbol{Q}}. \quad (51)$$

With (45) in (34), (50) $$\boldsymbol{F}^q = \boldsymbol{BD}^q \Rightarrow \boldsymbol{B} = (\boldsymbol{D}^q)^{-1} \Rightarrow \boldsymbol{B} = \sqrt[p^2]{\boldsymbol{Q}^q}. \quad (52)$$

With the above results, the $p$-valued Deutsch quantum gate with conjunctive control is shown in Fig. 8. Since the E and F blocks would realize only the identity (either activated or inhibited), they are omitted. Also the additional controlled-$\boldsymbol{X}$ gates at the output of the blocks, meant to recover the control states are not needed.

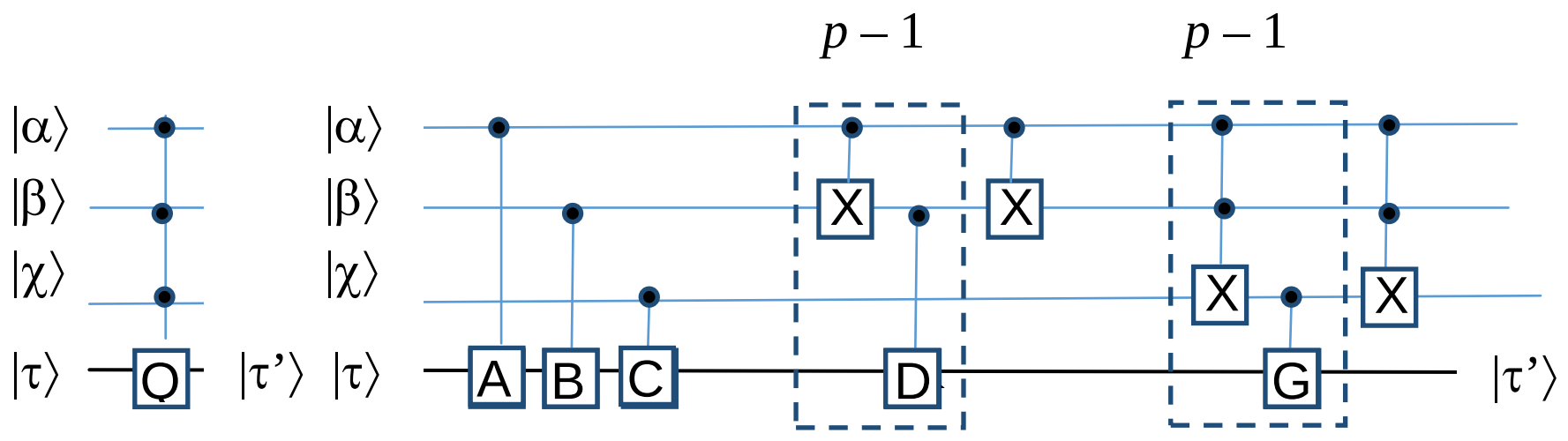

| A | B | C | D | G |
|---|---|---|---|---|
| $\sqrt[p^2]{\boldsymbol{Q}}$ | $\sqrt[p^2]{\boldsymbol{Q}^{p-1}}$ | $\sqrt[p]{\boldsymbol{Q}^{p-1}}$ | $\sqrt[p^2]{\boldsymbol{Q}^{-1}}$ | $\sqrt[p]{\boldsymbol{Q}^{-1}}$ |

Fig. 8 Realization of a $p$-valued Deutsch quantum gate with conjunctive control of three qupits.

| Table 5 : Cost analysis of the gate in Fig. 8 | |
|---|---|
| Initial gates with labels A, B, and C | 3 |
| Block D | $2(p-1)$ |
| Block G | $((2p+1)+1)(p-1) = 2(p^2-1)$ |
| Controlled-$\boldsymbol{X}$ gate to recover \|β⟩ | 1 |
| Controlled-controlled-$\boldsymbol{X}$ to recover \|χ⟩ | $2p+1$ |
| Total number of MS-gates | $2p^2 + 4p + 1 \in O(p^2)$ |

Notice that if the Deutsch gate is used to realize a quantum Toffoli gate, then $\boldsymbol{Q} = \boldsymbol{N}$ and the MS-gates A, D and G become $\boldsymbol{N}$, and B, C become $\boldsymbol{I}$. The total number of MS-gates will be reduced to $2p^2 + 4p - 1$. This is not only valid for a *CCCN* Toffoli gate, but for any *CCCQ* gate if $\boldsymbol{Q}$ is unitary and self inverse; for instance $\boldsymbol{Q} = \boldsymbol{XZ}^k\boldsymbol{N}$, where $\boldsymbol{Z} = Diag(1, \xi, \xi^2, \ldots, \xi^{p-1})$, $\xi$ is a primitive $p$-th root of unity, and $1 \le k \le p-1$. Besides $\boldsymbol{X}$, $\boldsymbol{Z}$ is also a Pauli matrix [17]. Any Hermitian matrix which is unitary, is self inverse. Permutation matrices which exchange independent pairs of elements of a vector, are also self inverse.

**Definition**: The Deutsch *class* comprises all *CCC**Q*** gates such that ***Q*** is unitary but not self inverse. The Toffoli *class* comprises all *CCC**Q*** gates where ***Q*** is both unitary and self inverse.

ii) Deduction of the elementary gates for the disjunctive control of gates of the Deutsch class.

The gate should be active, i.e. $|\tau'\rangle = \boldsymbol{Q}\,|\tau\rangle$, whenever at least one of the control qupits is in state $|p-1\rangle$.

Notice that Eqs. (34) and (36) are also valid for this case.

From row 3 and (36) $\quad \boldsymbol{ADE} = \boldsymbol{Q} \;\Rightarrow\; (\boldsymbol{AD})\boldsymbol{E} = \boldsymbol{E}^q\boldsymbol{E} = \boldsymbol{E}^p = \boldsymbol{Q} \;\Rightarrow\; \boldsymbol{E} = \sqrt[p]{\boldsymbol{Q}}.$ (53)

From row 2 and (34) $\quad \boldsymbol{BD}^q\boldsymbol{F} = \boldsymbol{Q} \;\Rightarrow\; (\boldsymbol{BD}^q)\boldsymbol{F} = \boldsymbol{F}^q\boldsymbol{F} = \boldsymbol{F}^p = \boldsymbol{Q} \;\Rightarrow\; \boldsymbol{F} = \sqrt[p]{\boldsymbol{Q}}.$ (54)

From row 5 and (36) $\quad \boldsymbol{ACDF}^q\boldsymbol{G}^q = \boldsymbol{Q} \;\Rightarrow\; \boldsymbol{CE}^q\boldsymbol{F}^q\boldsymbol{G}^q = \boldsymbol{Q} \;\Rightarrow\; \boldsymbol{E}^q\boldsymbol{F}^q\boldsymbol{G}^q = \boldsymbol{QC}^{-1}.$ (55)

From row 6 and (55) $\quad \boldsymbol{A}^q\boldsymbol{B}^q\boldsymbol{E}^q\boldsymbol{F}^q\boldsymbol{G}^q = \boldsymbol{Q}^q \;\Rightarrow\; \boldsymbol{A}^q\boldsymbol{B}^q\boldsymbol{QC}^{-1} = \boldsymbol{Q}^q \;\Rightarrow\; \boldsymbol{A}^q\boldsymbol{B}^q = \boldsymbol{CQ}^{p-2}.$ (56)

From row 7 and (56) $\quad \boldsymbol{ABC} = \boldsymbol{Q} \;\Rightarrow\; \boldsymbol{AB} = \boldsymbol{QC}^{-1} \;\Rightarrow\; \boldsymbol{A}^q\boldsymbol{B}^q = \boldsymbol{Q}^q(\boldsymbol{C}^{-1})^q \;\Rightarrow$

$\Rightarrow\; \boldsymbol{CQ}^{p-2} = \boldsymbol{Q}^q(\boldsymbol{C}^{-1}) \;\Rightarrow \boldsymbol{Q}^{-1} = (\boldsymbol{C}^{-1})(\boldsymbol{C}^{-1})^q = (\boldsymbol{C}^{-1})^p \;\Rightarrow$

$\Rightarrow\; \boldsymbol{C} = \sqrt[p]{\boldsymbol{Q}}.$ (57)

Notice that $\quad \boldsymbol{E} = \boldsymbol{F} = \boldsymbol{C}.$ (58)

From (55) and (58) $\quad \boldsymbol{E}^q\boldsymbol{F}^q\boldsymbol{G}^q = \boldsymbol{QC}^{-1} = (\boldsymbol{EF})^q\boldsymbol{G}^q = \boldsymbol{C}^{2q}\boldsymbol{G}^q \;\Rightarrow\; \boldsymbol{G}^q = \boldsymbol{QC}^{-1-2q} \;\Rightarrow$

$\Rightarrow\; \boldsymbol{G}^q = \boldsymbol{Q}\sqrt[p]{(\boldsymbol{Q})^{-1-2q}} = \sqrt[p]{(\boldsymbol{Q})^{-1-2q+p}} =$

$= \sqrt[p]{(\boldsymbol{Q})^{-(p-1)}} \;\Rightarrow\; \boldsymbol{G} = \sqrt[p]{\boldsymbol{Q}^{-1}}.$ (59)

From (53), (54), (59) $\quad \boldsymbol{EG} = \boldsymbol{I}$ and $\boldsymbol{FG} = \boldsymbol{I}.$ (60)

From rows 5 and 7 $\quad \boldsymbol{ACDF}^q\boldsymbol{G}^q = \boldsymbol{ACD} = \boldsymbol{Q} = \boldsymbol{ABC} \;\Rightarrow\; \boldsymbol{B} = \boldsymbol{D}.$ (61)

From rows 4 and 6 and with (58), (60), and (61) $\quad \boldsymbol{BCD}^q\boldsymbol{E}^q\boldsymbol{G}^q = \boldsymbol{ABEFG} = \boldsymbol{Q},$

$\boldsymbol{CD}^q = \boldsymbol{AF} \;\Rightarrow\; \boldsymbol{D}^q\boldsymbol{A}^{-1} = \boldsymbol{FC}^{-1} = \boldsymbol{I} \;\Rightarrow\; \boldsymbol{A} = \boldsymbol{D}^q.$ (62)

From row 3 and (62) $\quad \boldsymbol{ADE} = \boldsymbol{Q} \;\Rightarrow\; \boldsymbol{D}^p\boldsymbol{E} = \boldsymbol{Q} \;\Rightarrow\; \boldsymbol{D}^p = \boldsymbol{QE}^{-1} = \boldsymbol{Q}\sqrt[p]{\boldsymbol{Q}^{-1}} = \sqrt[p]{\boldsymbol{Q}^{p-1}} \Rightarrow$

With (61) $\quad \Rightarrow\; \boldsymbol{D} = \sqrt[p^2]{\boldsymbol{Q}^{p-1}} = \boldsymbol{B}.$ (63)

$\Rightarrow\; \boldsymbol{A} = \sqrt[p^2]{\boldsymbol{Q}^{(p-1)^2}}.$ (64)

The final gate for members of the Deutsch class has the structure shown in Fig. 6, with

| A | B | C | D | E | F | G |
|---|---|---|---|---|---|---|
| $\sqrt[p^2]{\boldsymbol{Q}^{(p-1)^2}}$ | $\sqrt[p^2]{\boldsymbol{Q}^{p-1}}$ | $\sqrt[p]{\boldsymbol{Q}}$ | $\sqrt[p^2]{\boldsymbol{Q}^{p-1}}$ | $\sqrt[p]{\boldsymbol{Q}}$ | $\sqrt[p]{\boldsymbol{Q}}$ | $\sqrt[p]{\boldsymbol{Q}^{-1}}$ |

It is simple to see that the cost of this gate will be that of Table 5, increased by the two simple blocks E and F, and the corresponding recovery controlled-***X*** gates, leading to

$$2p^2 + 4p + 1 + 2(2(p-1) + 1) = 2p^2 + 8p - 1 \text{ MS-gates} \qquad (65)$$

If the structure is used to realize gates of the Toffoli class with disjunctive control, the MS-gates with labels A, B, and D become ***I*** and those with labels C, E, F and G =will be ***Q***. The realization would require $2p^2 + 8p - 4$ MS-gates.

iii) Cost comparison of the ancillary-free realization of Deutsch gates with three control qubits and realizations based on decomposition and one ancillary line.

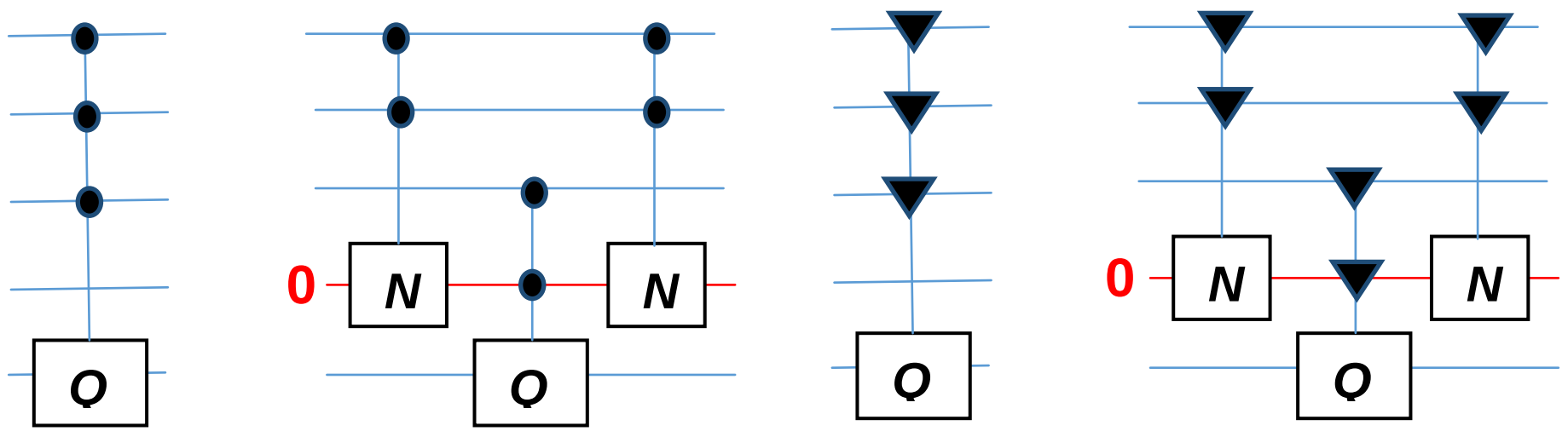

| $p$ | $2p^2 + 4p + 1$ | $6p + 1$ | $2p^2 + 8p – 1$ | $6p + 1$ |
|---|---|---|---|---|
| 3 | 31 | 19 | 41 | 19 |
| 5 | 71 | 31 | 89 | 31 |

Fig. 9: Comparison of the ancillary-free realization of *CCCQ* with a realization using *CCN*, *CCQ* and an ancillary line driven by 0.

iv) Mixed polarity in conjunctive control

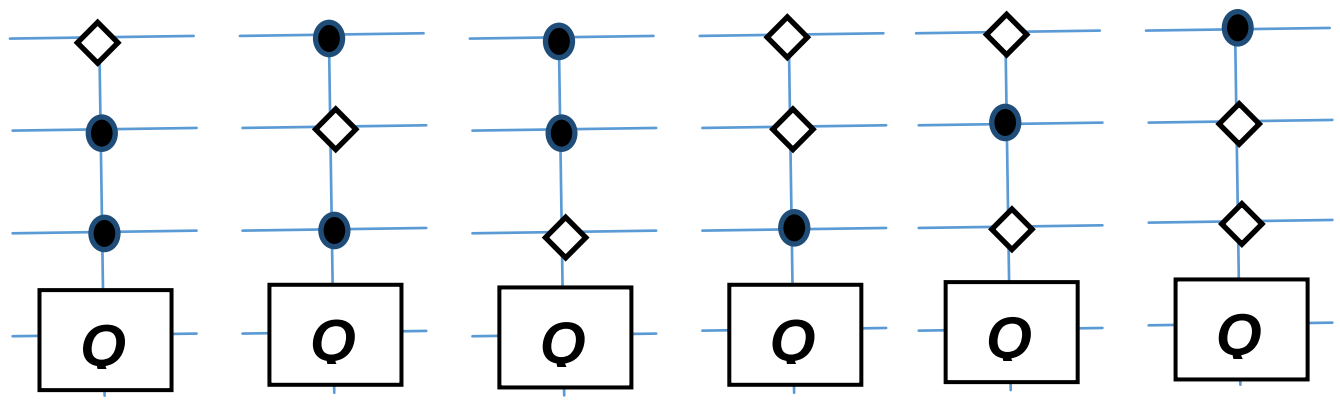

Case: 1 2 3 4 5 6

Fig. 10: Cases of mixed polarity on three control qupits.

The behavior of the gate with the different polarities is summarized in Table 6. Notice that the performance entries in rows 1, 3, and 5 is common to the cases 1, 3, and 5. (The repetition of numbers is just a coincidence). Similarly, the performance entries in rows 2 and 4, are common to the cases 2, 3, 4, and 6. Therefore, conclusions based on calculations on these entries will be valid for all the corresponding cases

Table 6. Abstract level behavior of gates of the Deutsch class ( Fig. 6) with mixed polarity

| | $\|\alpha\rangle$ | $\|\beta\rangle$ | $\|\chi\rangle$ | ***A*** | ***B*** | ***C*** | ***D*** | ***E*** | ***F*** | ***G*** | Case | | | | | |
|---|---|---|---|---|---|---|---|---|---|---|---|---|---|---|---|---|
| | | | | | | | | | | | 1 | 2 | 3 | 4 | 5 | 6 |
| 0 | $\|\omega\rangle$ | $\|\omega\rangle$ | $\|\omega\rangle$ | $I$ | $I$ | $I$ | $I^q$ | $I^q$ | $I^q$ | $I^q$ | $I$ | $I$ | $I$ | $I$ | $I$ | $I$ |
| 1 | $\|\omega\rangle$ | $\|\omega\rangle$ | $\|p–1\rangle$ | $I$ | $I$ | $C$ | $I^q$ | $E^q$ | $F^q$ | $G^q$ | $I$ | $I$ | $I$ | $Q$ | $I$ | $I$ |
| 2 | $\|\omega\rangle$ | $\|p–1\rangle$ | $\|\omega\rangle$ | $I$ | $B$ | $I$ | $D^q$ | $I^q$ | $F$ | $I^q$ | $I$ | $I$ | $I$ | $I$ | $Q$ | $I$ |
| 3 | $\|p–1\rangle$ | $\|\omega\rangle$ | $\|\omega\rangle$ | $A$ | $I$ | $I$ | $D$ | $E$ | $I^q$ | $I^q$ | $I$ | $I$ | $I$ | $I$ | $I$ | $Q$ |
| 4 | $\|\omega\rangle$ | $\|p–1\rangle$ | $\|p–1\rangle$ | $I$ | $B$ | $C$ | $D^q$ | $E^q$ | $I^q$ | $G^q$ | $Q$ | $I$ | $I$ | $I$ | $I$ | $I$ |
| 5 | $\|p–1\rangle$ | $\|\omega\rangle$ | $\|p–1\rangle$ | $A$ | $I$ | $C$ | $D$ | $I^q$ | $F^q$ | $G^q$ | $I$ | $Q$ | $I$ | $I$ | $I$ | $I$ |
| 6 | $\|p–1\rangle$ | $\|p–1\rangle$ | $\|\omega\rangle$ | $A$ | $B$ | $I$ | $I^q$ | $E$ | $F$ | $G$ | $I$ | $I$ | $Q$ | $I$ | $I$ | $I$ |
| 7 | $\|p–1\rangle$ | $\|p–1\rangle$ | $\|p–1\rangle$ | $A$ | $B$ | $C$ | $I^q$ | $I^q$ | $I^q$ | $I^q$ | $I$ | $I$ | $I$ | $I$ | $I$ | $I$ |

Case 1:

| | | |
|---|---|---|
| From rows 1, and 5 | $CE^qF^qG^q = ACDF^qG^q \Rightarrow AD = E^q.$ | (66) |
| From row 3 | $AD = E^{-1} \Rightarrow E^q = E^{-1} \Rightarrow E = I \Rightarrow AD = I \Rightarrow A^{-1} = D,$ | (67) |
| | $\Rightarrow F^qG^q = C^{-1}.$ | (68) |
| From rows 6 and 7 With (67) and (68) | $ABEFG = ABC = I \Rightarrow C = FG \Rightarrow C^q = F^qG^q \Rightarrow C^q = C^{-1},$ | |
| | $\Rightarrow C = I \Rightarrow F = G^{-1}.$ | (69) |
| From row 2 and (69) | $BD^qF = I \Rightarrow BD^q = F^{-1} = G\ .$ | (70) |
| From row 4 and (70) | $BCD^qG^q = Q \Rightarrow F^{-1}G^q = G^p = Q \Rightarrow G = \sqrt[p]{Q}.$ | (71) |
| | $\Rightarrow F = \sqrt[p]{Q^{-1}}.$ | (72) |
| From row 7 and (71) | $BC = A^{-1} = D \Rightarrow D^pG^q = Q \Rightarrow D^p = Q\sqrt[p]{Q^{-q}} = \sqrt[p]{Q^{p-q}} = \sqrt[p]{Q}$ | |
| | $\Rightarrow D = \sqrt[p^2]{Q} \Rightarrow A = \sqrt[p^2]{Q^{-1}}.$ | (73) |
| (73), (69) | $B = \sqrt[p^2]{Q}.$ | (74) |

The final gate for the Deutsch class has the structure shown in Fig.6, with

| A | B | C | D | E | F | G |
|---|---|---|---|---|---|---|
| $\sqrt[p^2]{Q^{-1}}$ | $\sqrt[p^2]{Q}$ | $I$ | $\sqrt[p^2]{Q}$ | $I$ | $\sqrt[p]{Q^{-1}}$ | $\sqrt[p]{Q}$ |

If the gate should belong to the Toffoli class, then all non-identity gates on the target line will be $CQ$.

Case 2:

| | | |
|---|---|---|
| From row 1 | $CE^qF^qG^q = I \Rightarrow C^{-1} = E^qF^qG^q.$ | (75) |
| From row 2 | $BD^qF = I \Rightarrow BD^q = F^{-1}.$ | (76) |
| From row 3 | $ADE = I \Longrightarrow AD = E^{-1}.$ | (77) |
| From rows 6 and 7 | $ABEFG = I = ABC \Longrightarrow C = EFG \Longrightarrow C^q = (EFG)^q.$ | (78) |
| From (78) and (75) | $C^{-1} = C^q \Longrightarrow C^p = I \Longrightarrow C = I,\ AB = I,\ EFG = I\ .$ | (79) |
| From row 5, (77), (79) | $ACDF^qG^q = Q \Longrightarrow AD(EFG)^q = E^qQ \Longrightarrow AD = E^qQ \Longrightarrow$ | |
| | $\Longrightarrow E^{-1} = E^qQ \Longrightarrow E^p = Q^{-1} \Longrightarrow E = \sqrt[p]{Q^{-1}}\ .$ | (80) |
| From row 4, (76), (79) | $BC(DEG)^q = B(DEG)^q = I \Longrightarrow B(D^q)(EG)^q = F^{-1}(EG)^q = I$ | |
| | $\Longrightarrow (EG)^q = F \Longrightarrow (EFG)^q = F^p = I \Longrightarrow F = I\ .$ | (81) |
| From (81) | $(EG)^q = F = I \Longrightarrow EG = I \Longrightarrow G = E^{-1} \Longrightarrow G = \sqrt[p]{Q}\ .$ | (82) |
| With (76), (77), (81) | $BD^q = I \Longrightarrow D^q = B^{-1} = A.$ | |
| | $AD = E^{-1} \Longrightarrow AD^p = E^{-1}D^q = E^{-1}A \Longrightarrow D^p = E^{-1} \Longrightarrow$ | |
| | $D = \sqrt[p]{E^{-1}} \Longrightarrow D = \sqrt[p^2]{Q}.$ | (83) |
| With (77), (80), (83) | $AD = E^{-1} \Longrightarrow A = D^{-1}E^{-1} = \sqrt[p^2]{Q^{-1}}\sqrt[p]{Q} = \sqrt[p^2]{Q^{-1}}\sqrt[p^2]{Q^p} \Longrightarrow$ | |
| | $\Longrightarrow A = \sqrt[p^2]{Q^{p-1}}.$ | (84) |
| With (79) and (84) | $B = \sqrt[p^2]{Q^{1-p}}.$ | (85) |

The final gate for the Deutsch class has the structure shown in Fig. 6, with

| A | B | C | D | E | F | G |
|---|---|---|---|---|---|---|
| $\sqrt[p^2]{Q^{p-1}}$ | $\sqrt[p^2]{Q^{1-p}}$ | $I$ | $\sqrt[p^2]{Q}$ | $\sqrt[p]{Q^{-1}}$ | $I$ | $\sqrt[p]{Q}$ |

If the gate should belong to the Toffoli class, then the gates labeled A and B will be $\boldsymbol{I}$ (and can be removed), and those labeled D, E and G will be $\boldsymbol{Q}$ thus requiring just $3(p-1)$ $C\boldsymbol{Q}$ gates on the target line.

Case 3:

From Case 1 $\quad \boldsymbol{E = I}, \boldsymbol{A = D^{-1}}, \ (\boldsymbol{FG})^q = \boldsymbol{C}^{-1} = \boldsymbol{F}^q\boldsymbol{G}^q.$ (86)

From rows 2 , 4, (86) $\quad \boldsymbol{BD^qF = BCD^qE^qG^q} \implies \boldsymbol{F = CG^q} = (\boldsymbol{F}^{-q}\boldsymbol{G}^{-q})\boldsymbol{G}^q \implies \boldsymbol{F = F^{-q}}$
$\implies \boldsymbol{F = I} \implies \boldsymbol{B = D^{-q}}, \boldsymbol{C = G^{-q}}.$ (87)

From (86) and (87) $\quad \boldsymbol{AB = D^{-1}D^{-q} = D^{-p}}.$ (88)

From row 7 $\quad \boldsymbol{AB = C^{-1}}.$ (89)

From row 6, (87), (88) $\quad \boldsymbol{ABEFG = ABG = Q} \implies \boldsymbol{D^{-p}G = Q}$ and $\boldsymbol{C^{-1}G = Q} \implies \boldsymbol{G^p = Q}$
$\boldsymbol{G} = \sqrt[p]{\boldsymbol{Q}} \implies \boldsymbol{C} = \sqrt[p]{\boldsymbol{Q}^{-q}}.$ (90)

From (90) and (86) $\quad \boldsymbol{D^{-p}G = Q} \implies \boldsymbol{D}^p = \boldsymbol{Q}^{-1}\boldsymbol{G} = \boldsymbol{Q}^{-1}\sqrt[p]{\boldsymbol{Q}} = \sqrt[p]{\boldsymbol{Q}^{-q}} \implies$
$\implies \boldsymbol{D} = \sqrt[p^2]{\boldsymbol{Q}^{-q}}$ and $\boldsymbol{A} = \sqrt[p^2]{\boldsymbol{Q}^{q}}$. (91)

From (89), (90), (91) $\quad \boldsymbol{B} = \boldsymbol{A}^{-1}\boldsymbol{C}^{-1} = \sqrt[p^2]{\boldsymbol{Q}^{-q}}\sqrt[p]{\boldsymbol{Q}^{q}} = \sqrt[p^2]{\boldsymbol{Q}^{-q}}\sqrt[p^2]{\boldsymbol{Q}^{pq}} = \sqrt[p^2]{\boldsymbol{Q}^{q(p-1)}} \implies$
$\boldsymbol{B} = \sqrt[p^2]{\boldsymbol{Q}^{q^2}}$. (92)

The final gate for the Deutsch class has the structure shown in Fig. 6, with

| A | B | C | D | E | F | G |
|---|---|---|---|---|---|---|
| $\sqrt[p^2]{Q^{q}}$ | $\sqrt[p^2]{Q^{q^2}}$ | $\sqrt[p]{Q^{-q}}$ | $\sqrt[p^2]{Q^{-q}}$ | $I$ | $I$ | $\sqrt[p]{Q}$ |

Notice that in this case, gates of the Toffoli class are simply realized with the block G, with G corresponding to $\boldsymbol{Q}$, since the $CCX$ controlling gates driven by $|\alpha\rangle$ and $|\beta\rangle$ will be active when these qupits are in state $|p-1\rangle$ and for $|\chi\rangle \in \Omega$, exactly one of the G gates will be activated, meanwhile if $|\chi\rangle = |p-1\rangle$ this qupit will be shifted through all states of $\Omega$ without reaching again the state $|p-1\rangle$ (inside the block) leaving the G gates inhibited.

Case 4:

From row 1 $\quad \boldsymbol{CE^qF^qG^q = Q} \implies \boldsymbol{CE^qG^q = QF^{-q}}$ and $\boldsymbol{CF^qG^q = QE^{-q}}.$ (93)

From rows 2, 4, (93) $\quad \boldsymbol{BD^qF = BCD^qE^qG^q} \implies \boldsymbol{F = CE^qG^q} \implies \boldsymbol{F = QF^{-q}} \implies \boldsymbol{F} = \sqrt[p]{\boldsymbol{Q}}.$ (94)
$\boldsymbol{BD^qF = I} \implies \boldsymbol{BD^q = F^{-1}}.$ (95)

From rows 3 and 5 $\quad \boldsymbol{ADE = ACDF^qG^q} \implies \boldsymbol{CF^qG^q = E = QE^{-q}} \implies \boldsymbol{E} = \sqrt[p]{\boldsymbol{Q}} = \boldsymbol{F}.$ (96)

From rows 6, 7, (93) $\quad \boldsymbol{ABEFG = ABC} \implies \boldsymbol{C = EFG} \implies \boldsymbol{C^q = E^qF^qG^q} \implies \boldsymbol{C^p = Q} \implies$
$\implies \boldsymbol{C} = \sqrt[p]{\boldsymbol{Q}} = \boldsymbol{E} = \boldsymbol{F}.$ (97)

From (94), (97) $\quad \boldsymbol{F = C = CE^qG^q} \implies \boldsymbol{E^q = G^{-q}} \implies \boldsymbol{G = E^{-1}} \implies \boldsymbol{G} = \sqrt[p]{\boldsymbol{Q}^{-1}}.$ (98)

From row 7, (96), (98) $\quad \boldsymbol{ADE = I = ABC} \implies \boldsymbol{D = B}.$ (99)

From (95), (99) $BD^q = F^{-1} \Rightarrow D^p = F^{-1} \Rightarrow D = \sqrt[p]{F^{-1}} = \sqrt[p^2]{Q^{-1}} = B.$ (100)

From row 3, (96), (100) $A = D^{-1}E^{-1} = \sqrt[p^2]{Q}\,\sqrt[p]{Q^{-1}} = \sqrt[p^2]{Q^1}\,\sqrt[p^2]{Q^{-p}} = \sqrt[p^2]{Q^{1-p}} = A.$ (101)

The final gate has the structure shown in Fig. 6, with

| A | B | C | D | E | F | G |
|---|---|---|---|---|---|---|
| $\sqrt[p^2]{Q^{1-p}}$ | $\sqrt[p^2]{Q^{-1}}$ | $\sqrt[p]{Q}$ | $\sqrt[p^2]{Q^{-1}}$ | $\sqrt[p]{Q}$ | $\sqrt[p]{Q}$ | $\sqrt[p]{Q^{-1}}$ |

For the realization of gates of the Toffoli class, A will become ***I*** and all others, ***Q.***

Case 5:

From row 1 $CE^qF^qG^q = I \Rightarrow CF^qG^q = E^{-q}.$ (102)

From rows 3, 5, (102) $ACDF^qG^q \Rightarrow ADE^{-q} = I = ADE \Rightarrow E = E^{-q} \Rightarrow E = I.$ (103)

$\Rightarrow A = D^{-1};\ CF^qG^q = I \Rightarrow C = (F^qG^q)^{-1}.$ (104)

From rows 6, 7, (104) $AB(E)FG = ABC \Rightarrow C = FG = (F^qG^q)^{-1} \Rightarrow C = I.$ (105)

$\Rightarrow A = B^{-1};\ B = D;\ F = G^{-1}.$ (106)

From rows 2, 4 $B(C)D^q(E^q)G^q = I \Rightarrow BD^q = (G^q)^{-1};\ BD^qF = Q \Rightarrow$ (107)

$\Rightarrow (G^q)^{-1}F = Q \Rightarrow F^qF = F^p = Q \Rightarrow F = \sqrt[p]{Q} \Rightarrow G = \sqrt[p]{Q^{-1}}.$ (108)

From (107), (108), (106), (104) $BD^q = (G^q)^{-1} \Rightarrow DD^q = D^p = (G^q)^{-1} = \sqrt[p]{Q^q} \Rightarrow$

$\Rightarrow D = B = \sqrt[p^2]{Q^{p-1}} \Rightarrow A = \sqrt[p^2]{Q^{1-p}}.$ (109)

The final gate for the Deutsch class has the structure shown in Fig. 6, with

| A | B | C | D | E | F | G |
|---|---|---|---|---|---|---|
| $\sqrt[p^2]{Q^{1-p}}$ | $\sqrt[p^2]{Q^{p-1}}$ | $I$ | $\sqrt[p^2]{Q^{p-1}}$ | $I$ | $\sqrt[p]{Q}$ | $\sqrt[p]{Q^{-1}}$ |

For the realization of gates of the Toffoli class, F and G represent ***Q*** gates. All others will become ***I*** and may be removed.

Case 6:

From row 1 $CE^qF^qG^q = I \Rightarrow CE^qG^q = F^{-q};\ CF^qG^q = E^{-q}.$ (110)

From row 2 $BD^qF = I \Rightarrow BD^q = F^{-1}.$ (111)

With 1 and 2 in 4 $BCD^qE^qG^q = (BD^q)(CE^qG^q) = F^{-1}F^{-q} = I \Rightarrow F = I.$ (112)

Row 5 with 1 and 3 $ACDF^qG^q = ADE^{-q} = I \Rightarrow QE^{-1}E^{-q} = QE^{-p} = I \Rightarrow E = \sqrt[p]{Q}.$ (113)

Rows 6, 7 with 1 $ABEFG = ABC \Rightarrow C = EFG \Rightarrow CE^qF^qG^q = (EG)^p = I \Rightarrow$ (114)

$\Rightarrow EG = I \Rightarrow G = \sqrt[p]{Q^{-1}}.$ (115)

From (114) and (110) $C = EFG = I \Rightarrow C^q = E^qF^qG^q \Rightarrow C(E^qF^qG^q) = I \Rightarrow C^p = I \Rightarrow C = I.$ (116)

From 2 with (112) $BD^q = I \Rightarrow B = D^{-q}.$ (117)

From 7 with (116) $ABC = I \Rightarrow AB = I \Rightarrow A = B^{-1} = D^q.$ (118)

From 5 with (118) $ACDF^qG^q = ADG^q = D^qDG^q = D^pG^q = I \Rightarrow D = \sqrt[p]{G^{-q}} \Rightarrow$

$\Rightarrow D = \sqrt[p^2]{Q^q} \Rightarrow A = \sqrt[p^2]{Q^{q^2}} \Rightarrow B = \sqrt[p^2]{Q^{-q^2}}.$ (119)

The final gate for the Deutsch class has the structure shown in Fig. 6, with

| A | B | C | D | E | F | G |
|---|---|---|---|---|---|---|
| $\sqrt[p^2]{\boldsymbol{Q}^{(p-1)^2}}$ | $\sqrt[p^2]{\boldsymbol{Q}^{-(p-1)^2}}$ | $\boldsymbol{I}$ | $\sqrt[p^2]{\boldsymbol{Q}^{p-1}}$ | $\sqrt[p]{\boldsymbol{Q}}$ | $\boldsymbol{I}$ | $\sqrt[p]{\boldsymbol{Q}^{-1}}$ |

For the realization of gates of the Toffoli class, E and G represent $\boldsymbol{Q}$ gates. All others will become $\boldsymbol{I}$ and (the corresponding blocks) may be removed.

Table 7: Summary of realizations of *p*-valued Deutsch gates and gates of the Toffoli class, with three control qupits and mixed polarity

| Case | | A | B | C | D | E | F | G |
|---|---|---|---|---|---|---|---|---|
| 1 | Deutsch class | $\sqrt[p^2]{\boldsymbol{Q}^{-1}}$ | $\sqrt[p^2]{\boldsymbol{Q}}$ | $\boldsymbol{I}$ | $\sqrt[p^2]{\boldsymbol{Q}}$ | $\boldsymbol{I}$ | $\sqrt[p]{\boldsymbol{Q}^{-1}}$ | $\sqrt[p]{\boldsymbol{Q}}$ |
| | Toffoli class | $\boldsymbol{Q}$ | $\boldsymbol{Q}$ | $\boldsymbol{I}$ | $\boldsymbol{Q}$ | $\boldsymbol{I}$ | $\boldsymbol{Q}$ | $\boldsymbol{Q}$ |
| 2 | Deutsch class | $\sqrt[p^2]{\boldsymbol{Q}^{p-1}}$ | $\sqrt[p^2]{\boldsymbol{Q}^{1-p}}$ | $\boldsymbol{I}$ | $\sqrt[p^2]{\boldsymbol{Q}}$ | $\sqrt[p]{\boldsymbol{Q}^{-1}}$ | $\boldsymbol{I}$ | $\sqrt[p]{\boldsymbol{Q}}$ |
| | Toffoli class | $\boldsymbol{I}$ | $\boldsymbol{I}$ | $\boldsymbol{I}$ | $\boldsymbol{Q}$ | $\boldsymbol{Q}$ | $\boldsymbol{I}$ | $\boldsymbol{Q}$ |
| 3 | Deutsch class | $\sqrt[p^2]{\boldsymbol{Q}^{p-1}}$ | $\sqrt[p^2]{\boldsymbol{Q}^{(p-1)^2}}$ | $\sqrt[p]{\boldsymbol{Q}^{1-p}}$ | $\sqrt[p^2]{\boldsymbol{Q}^{1-p}}$ | $\boldsymbol{I}$ | $\boldsymbol{I}$ | $\sqrt[p]{\boldsymbol{Q}}$ |
| | Toffoli class | $\boldsymbol{I}$ | $\boldsymbol{I}$ | $\boldsymbol{I}$ | $\boldsymbol{I}$ | $\boldsymbol{I}$ | $\boldsymbol{I}$ | $\boldsymbol{Q}$ |
| 4 | Deutsch class | $\sqrt[p^2]{\boldsymbol{Q}^{1-p}}$ | $\sqrt[p^2]{\boldsymbol{Q}^{-1}}$ | $\sqrt[p]{\boldsymbol{Q}}$ | $\sqrt[p^2]{\boldsymbol{Q}^{-1}}$ | $\sqrt[p]{\boldsymbol{Q}}$ | $\sqrt[p]{\boldsymbol{Q}}$ | $\sqrt[p]{\boldsymbol{Q}^{-1}}$ |
| | Toffoli class | $\boldsymbol{I}$ | $\boldsymbol{Q}$ | $\boldsymbol{Q}$ | $\boldsymbol{Q}$ | $\boldsymbol{Q}$ | $\boldsymbol{Q}$ | $\boldsymbol{Q}$ |
| 5 | Deutsch class | $\sqrt[p^2]{\boldsymbol{Q}^{1-p}}$ | $\sqrt[p^2]{\boldsymbol{Q}^{p-1}}$ | $\boldsymbol{I}$ | $\sqrt[p^2]{\boldsymbol{Q}^{p-1}}$ | $\boldsymbol{I}$ | $\sqrt[p]{\boldsymbol{Q}}$ | $\sqrt[p]{\boldsymbol{Q}^{-1}}$ |
| | Toffoli class | $\boldsymbol{I}$ | $\boldsymbol{I}$ | $\boldsymbol{I}$ | $\boldsymbol{I}$ | $\boldsymbol{I}$ | $\boldsymbol{Q}$ | $\boldsymbol{Q}$ |
| 6 | Deutsch class | $\sqrt[p^2]{\boldsymbol{Q}^{(p-1)^2}}$ | $\sqrt[p^2]{\boldsymbol{Q}^{-(p-1)^2}}$ | $\boldsymbol{I}$ | $\sqrt[p^2]{\boldsymbol{Q}^{p-1}}$ | $\sqrt[p]{\boldsymbol{Q}}$ | $\boldsymbol{I}$ | $\sqrt[p]{\boldsymbol{Q}^{-1}}$ |
| | Toffoli class | $\boldsymbol{I}$ | $\boldsymbol{I}$ | $\boldsymbol{I}$ | $\boldsymbol{I}$ | $\boldsymbol{Q}$ | $\boldsymbol{I}$ | $\boldsymbol{Q}$ |

v) Special disjunctive controls with mixed polarity

Case 7: A gate is active iff *any* two control qupits are in state $|p-1\rangle$ and the state of the remaining control qupit is in Ω. Otherwise the gate is inhibited and behaves as the identity. (This represents the disjunctive union of Case 4 ***or*** Case 5 ***or*** Case 6).

Case 8: A gate is active iff *any* control qupit is in state $|p-1\rangle$ and the states of the remaining two control qupits are in Ω. Otherwise the gate is inhibited and behaves as the identity. (This represents the disjunctive union of Case 1 ***or*** Case 2 ***or*** Case 3).

The behavior is specified in Table 8.

| Table 8. Abstract level behavior of the gate with special disjunctive control and mixed polarity | | | | | | | | | | | | |
|---|---|---|---|---|---|---|---|---|---|---|---|---|
| | \|α⟩ | \|β⟩ | \|χ⟩ | ***A*** | ***B*** | ***C*** | ***D*** | ***E*** | ***F*** | ***G*** | Case | |
| | | | | | | | | | | | 7 | 8 |
| 0 | \|ω⟩ | \|ω⟩ | \|ω⟩ | $I$ | $I$ | $I$ | $I^q$ | $I^q$ | $I^q$ | $I^q$ | $I$ | $I$ |
| 1 | \|ω⟩ | \|ω⟩ | \|p–1⟩ | $I$ | $I$ | $C$ | $I^q$ | $E^q$ | $F^q$ | $G^q$ | $I$ | $Q$ |
| 2 | \|ω⟩ | \|p–1⟩ | \|ω⟩ | $I$ | $B$ | $I$ | $D^q$ | $I^q$ | $F$ | $I^q$ | $I$ | $Q$ |
| 3 | \|p–1⟩ | \|ω⟩ | \|ω⟩ | $A$ | $I$ | $I$ | $D$ | $E$ | $I^q$ | $I^q$ | $I$ | $Q$ |
| 4 | \|ω⟩ | \|p–1⟩ | \|p–1⟩ | $I$ | $B$ | $C$ | $D^q$ | $E^q$ | $I^q$ | $G^q$ | $Q$ | $I$ |
| 5 | \|p–1⟩ | \|ω⟩ | \|p–1⟩ | $A$ | $I$ | $C$ | $D$ | $I^q$ | $F^q$ | $G^q$ | $Q$ | $I$ |
| 6 | \|p–1⟩ | \|p–1⟩ | \|ω⟩ | $A$ | $B$ | $I$ | $I^q$ | $E$ | $F$ | $G$ | $Q$ | $I$ |
| 7 | \|p–1⟩ | \|p–1⟩ | \|p–1⟩ | $A$ | $B$ | $C$ | $I^q$ | $I^q$ | $I^q$ | $I^q$ | $I$ | $I$ |

Case 7:

From row 1
$$\boldsymbol{CE^qF^qG^q = I \implies C = (E^qF^qG^q)^{-1} \implies EFG = \sqrt[q]{C^{-1}},} \quad (120)$$
$$\boldsymbol{\implies CE^qG^q = (F^q)^{-1} \;;\; CF^qG^q = (E^q)^{-1}.} \quad (121)$$

From rows 6, 7, (120)
$$\boldsymbol{AB = C^{-1}\,; ABEFG = Q \implies C^{-1}EFG = C^{-1}\sqrt[q]{C^{-1}} = Q \implies}$$
$$\boldsymbol{\implies C\sqrt[q]{C} = Q^{-1} \implies \sqrt[q]{C^p} = Q^{-1} \implies C = \sqrt[p]{Q^{1-p}}.} \quad (122)$$

From 2, 4, (121)
$$\boldsymbol{BD^q = F^{-1}\,;\; BCD^qE^qG^q = BD^q(F^q)^{-1} = F^{-1}(F^q)^{-1} = Q \implies}$$
$$\boldsymbol{(F^p)^{-1} = Q \implies F = \sqrt[p]{Q^{-1}}.} \quad (123)$$

From rows 3, 5, (121)
$$\boldsymbol{AD = E^{-1}\,; ACDF^qG^q = E^{-1}CF^qG^q = E^{-1}(E^q)^{-1} = Q \implies}$$
$$\boldsymbol{(E^p)^{-1} = Q \implies E = \sqrt[p]{Q^{-1}}.} \quad (124)$$

From (120) to (124)
$$\boldsymbol{G = E^{-1}F^{-1}\sqrt[q]{C^{-1}} = \sqrt[p]{Q^2}\,\sqrt[pq]{Q^{p-1}} = \sqrt[p]{Q^2}\,\sqrt[p]{Q} \implies G = \sqrt[p]{Q^3}.} \quad (125)$$

Multiply rows 2 and 3
$$\boldsymbol{(BD^qF = I\,)(ADE = I) = ABD^pEF = I \implies ABEF = D^{-p}.} \quad (126)$$

From row 6 and (126)
$$\boldsymbol{ABEFG = Q \implies D^{-p}G = Q \implies D^{-p} = QG^{-1} = Q\sqrt[p]{Q^{-3}} =}$$
$$\boldsymbol{= \sqrt[p]{Q^{p-3}} \implies D = \sqrt[p^2]{Q^{3-p}}.} \quad (127)$$

From row 3, (127) and (124)
$$\boldsymbol{A = (DE)^{-1} = \left(\sqrt[p^2]{Q^{3-p}}\,\sqrt[p]{Q^{-1}}\right)^{-1} = \sqrt[p^2]{Q^{p-3}}\,\sqrt[p^2]{Q^p} = \sqrt[p^2]{Q^{2p-3}}.} \quad (128)$$

From row 7
$$\boldsymbol{B = (AC)^{-1} = \left(\sqrt[p^2]{Q^{2p-3}}\,\sqrt[p]{Q^{1-p}}\right)^{-1} = \sqrt[p^2]{Q^{p^2-3p+3}}.} \quad (129)$$

The final gate for the Deutsch class has the structure shown in Fig. 6, with

| A | B | C | D | E | F | G |
|---|---|---|---|---|---|---|
| $\sqrt[p^2]{Q^{2p-3}}$ | $\sqrt[p^2]{Q^{p^2-3p+3}}$ | $\sqrt[p]{Q^{1-p}}$ | $\sqrt[p^2]{Q^{3-p}}$ | $\sqrt[p]{Q^{-1}}$ | $\sqrt[p]{Q^{-1}}$ | $\sqrt[p]{Q^3}$ |

For the realization of gates of the Toffoli class, and C and D become ***I*** and all others realize ***Q***.

Case 8:

Rows 1, 6 and 7
$$\boldsymbol{ABC = ABEFG = I \implies C = EFG,} \quad (130)$$

$$CE^qF^qG^q = Q \Rightarrow C^p = Q \Rightarrow C = \sqrt[p]{Q}. \tag{131}$$

Rows 2 and 3
$$BD^qF = Q \Rightarrow BD^q = QF^{-1}. \tag{132}$$
$$ADE = Q \Rightarrow AD = QE^{-1}. \tag{133}$$

Rows 4, 5, (132-3)
$$BCD^qE^qG^q = ACDF^qG^q \Rightarrow ADF^q = BD^qE^q \Rightarrow$$
$$\Rightarrow QE^{-1}F^q = QF^{-1}E^q \Rightarrow F^p = E^p \Rightarrow E = F. \tag{134}$$

Rows 1 and 4, (132)
$$CE^qG^q = QF^{-q} \Rightarrow BD^qQF^{-q} = I \Rightarrow Q^2F^{-1}F^{-q} = I \Rightarrow F^p = Q^2$$
$$\Rightarrow F = E = \sqrt[p]{Q^2}. \tag{135}$$

Multiply 2, 3; 7, (134)
$$ABD^p = Q^2E^{-2} \Rightarrow D^p = CQ^2E^{-2} = \sqrt[p]{Q}Q^2\sqrt[p]{Q^{-4}} = \sqrt[p]{Q^{2p-3}} \Rightarrow$$
$$\Rightarrow D = \sqrt[p^2]{Q^{2p-3}}. \tag{136}$$

From (130), (134)
$$C = EFG = E^2G \Rightarrow G = CE^{-2} = \sqrt[p]{Q}\sqrt[p]{Q^{-4}} = \sqrt[p]{Q^{-3}} = G. \tag{137}$$

From (133)
$$AD = QE^{-1} = Q\sqrt[p]{Q^{-2}} = \sqrt[p]{Q^{p-2}},$$
$$A = QE^{-1}D^{-1} = \sqrt[p]{Q^{p-2}}\,\sqrt[p^2]{Q^{3-2p}} = \sqrt[p^2]{Q^{p(p-2)}}\,\sqrt[p^2]{Q^{3-2p}} \Rightarrow$$
$$\Rightarrow A = \sqrt[p^2]{Q^{p^2-4p+3}}. \tag{138}$$

From (132)
$$BD^q = QF^{-1} \Rightarrow B = QF^{-1}D^{1-p} = Q\sqrt[p]{Q^{-2}}\,\sqrt[p^2]{Q^{(2p-3)(p-1)}} \Rightarrow$$
$$\Rightarrow B = \sqrt[p^2]{Q^{-(p^2-3p+3)}}. \tag{139}$$

The final gate for the Deutsch class has the structure shown in Fig. 6, with

| A | B | C | D | E | F | G |
|---|---|---|---|---|---|---|
| $\sqrt[p^2]{Q^{p^2-4p+3}}$ | $\sqrt[p^2]{Q^{-(p^2-3p+3)}}$ | $\sqrt[p]{Q}$ | $\sqrt[p^2]{Q^{2p-3}}$ | $\sqrt[p]{Q^2}$ | $\sqrt[p]{Q^2}$ | $\sqrt[p]{Q^{-3}}$ |

For the realization of gates of the Toffoli class, the gates with labels A, E and F correspond to $\boldsymbol{Q}$, and gates with labels B, C, D and G correspond to $\boldsymbol{I}$, therefore the gates with labels B and C, and the blocks with labels D and G may be removed.

## VI ON THE COMPLEXITY OF REALIZATIONS FOR ANY NUMBER OF CONTROL QUPITS

In [2], for the realization of (binary) Toffoli gates with $n$ control signals, the gates on the target line were controlled by signals generated as $s_1c_1 \oplus s_2c_2 \oplus \ldots \oplus s_nc_n$, where for $1 \le i \le n$ the $c_i$ denoted the control signals, $s_i \in \{0, 1\}$, the selectors (of the control signals activating different auxiliary gates on the target line) and $\oplus$ meant the sum mod 2. In [2] it was shown that if the "words" $s_1s_2\ldots s_n$ were taken following a Grey code (without the word 00…0), the realization would have minimal cost. In [14] it was shown that the minimal cost was also possible if the selector words were ordered by increasing Hamming weight, and in [15], that this was also possible by ordering the selector words after a binary code with bit-reversal.

Let now the case of gates of the Deutsch class with three control qupits be analyzed. In analogy to [14] for the realization of the gate a "pseudo Hamming weight" will be considered, where the pseudo Hamming weight corresponds to the number of qupits that control a given sub-circuit or gate on the target line. A symbolic representation is straightforward when the pseudo Hamming weight is 1 or 2, but when the pseudo Hamming weight becomes 3 the symbolic representation would lead to a "recursive" kind of realization. Therefore in this case it should be specified that two qupits control an $\boldsymbol{X}$ gate, which will rotate the third qupit before driving a gate on the target line. This scheme is shown in Fig. 11, where the only predefined gates are the $\boldsymbol{X}$ gates.

$|\alpha\rangle$
$|\beta\rangle$
$|\gamma\rangle$

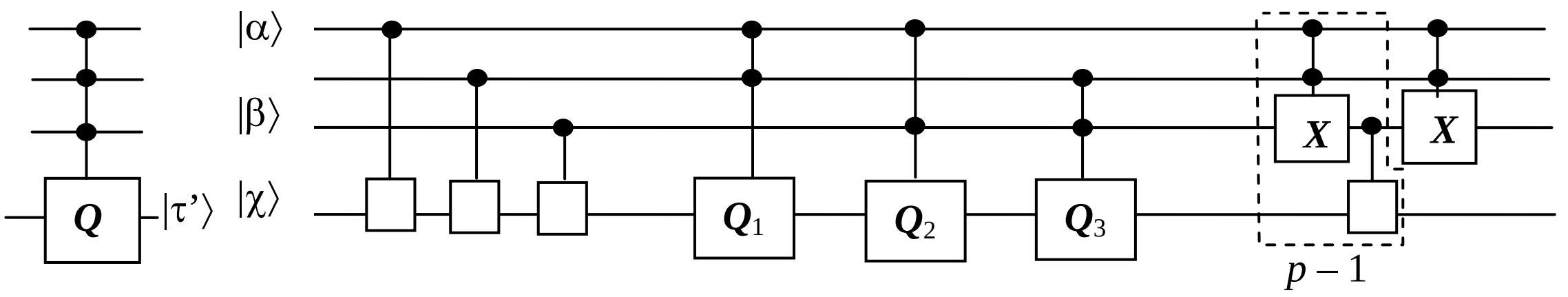

Fig. 11: Symbolic representation of a realization scheme for a Deutsch gate with three control qupits

The controlled-controlled $\boldsymbol{Q}_1$, $\boldsymbol{Q}_2$ and $\boldsymbol{Q}_3$ sub-circuits may be realized as shown in Fig. 4. If this is introduced in the above scheme, the structure shown in Fig. 12 is obtained.

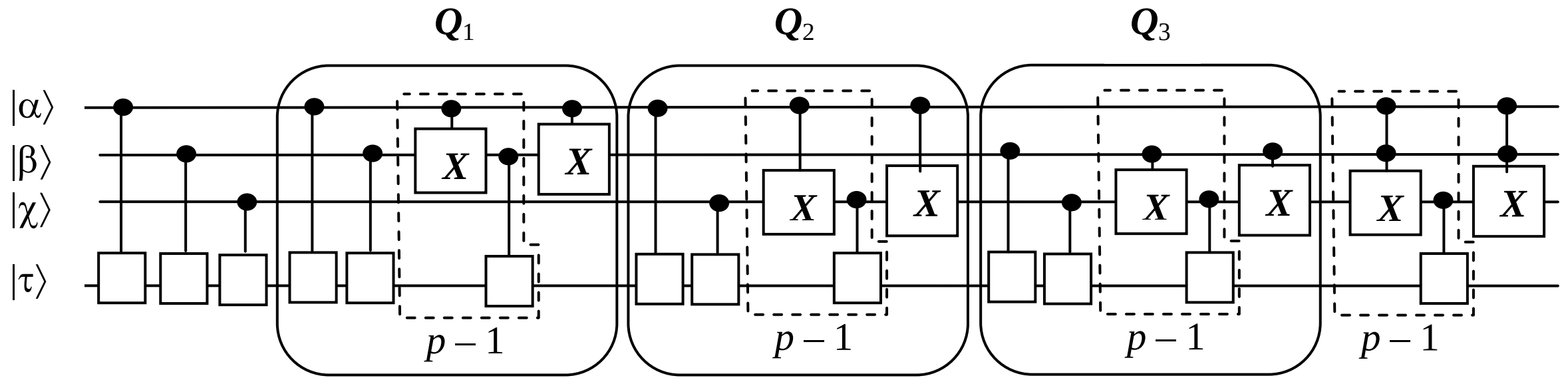

Fig. 12: Realization scheme for a Deutsch gate with three control qupits with explicit expansion of the controlled-controlled sub-circuits.

It should be noted that the controlled-controlled sub-circuits of Fig. 12 preserve the controlling conditions shown in Fig. 11, i.e., each sub-circuit is controlled-controlled by qupits in the same state as in the main input. Therefore, repeated single-controlled gates may be merged by multiplying the unitary matrices of their corresponding specifications. This leads to the final circuit shown in Fig. 13, which has the same structure as the initial conjectured circuit of Fig. 8. Depending on the kind of expected control (conjunctive or disjunctive) and the selected polarity, the elementary gates on the target line will have different specifications, as shown in the former sections.

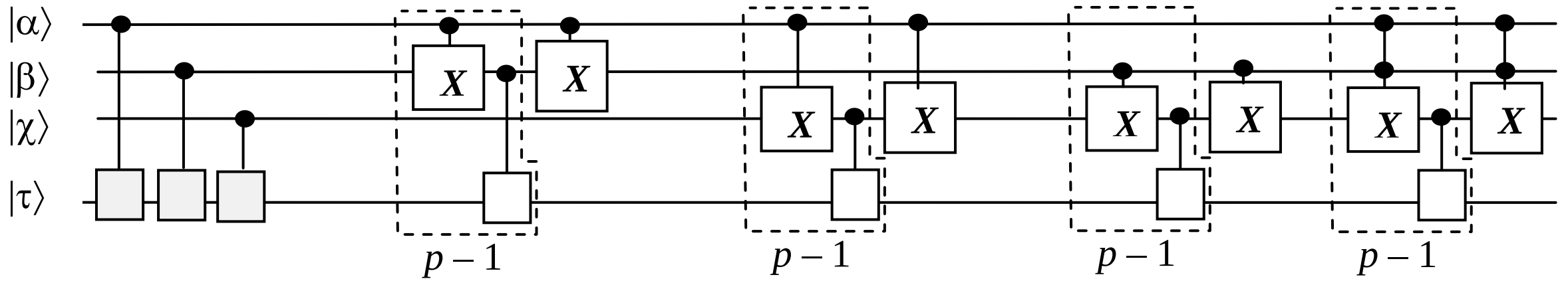

Fig. 13: Final realization scheme for a Deutsch gate with three controlling qupits.

The scheme of Fig. 13 shows that the first parts of the final gate, corresponding to the pseudo Hamming weights of 1 and 2, may be realized exclusively using elementary gates with single control, and the number of required gates in is $6p$. The last part, corresponding to the pseudo Hamming weight of 3, requires the use of $(p - 1)$ controlled-controlled $\boldsymbol{X}$ gates. These controlled-controlled-$\boldsymbol{X}$ gates may however be realized as shown in Fig. 12 for the sub-circuits $\boldsymbol{Q}_1$, $\boldsymbol{Q}_2$ and $\boldsymbol{Q}_3$, requiring again single control gates, but including a block that has to be repeated $(p - 1)$ times, thus reaching a realization complexity in $O(p^2)$. Inductive reasoning allows to see that for any $n$ controlling qupits, a realization with a block containing $(p - 1)$ $\boldsymbol{X}$ gates controlled by $(n - 1)$ qupits will be needed; where ach one of them may be realized by a sub-block containing $(p - 1)$ $\boldsymbol{X}$ gates controlled by $(n - 2)$ qupits. It becomes apparent that the realization complexity will be in $O(p^{n-1})$, where this represent the number of single controlled elementary gates needed for the realization. This, at the same time, illustrates the Muthukrishnan-Stroud realizability of these gates.